\documentclass[10pt,twocolumn]{article} 
\usepackage{latex8}

\usepackage{times,boxedminipage,epsfig,psfig,alltt,subfigure,amsmath}
\newcommand{\comment}[1]{}
\newcommand{\COMMENT}[1]{}
\newcommand{\jhcomment}[1]{}
\newcommand{\shcomment}[1]{}
\def\choose#1#2{ \binom{#1}{#2} }

\begin{document}
\title{A Framework for High-Accuracy Privacy-Preserving Mining\\
\vspace*{0.6in}
}
\author{Shipra Agrawal \qquad Jayant R. Haritsa\\
Database Systems Lab, SERC/CSA \\
Indian Institute of Science, Bangalore 560012, INDIA\\
\{shipra,haritsa\}@dsl.serc.iisc.ernet.in
}

\maketitle
\begin{abstract}
To preserve client privacy in the data mining process, a variety of
techniques based on random perturbation of data records have been proposed
recently.  In this paper, we present a generalized matrix-theoretic
model of random perturbation, which facilitates a systematic approach
to the design of perturbation mechanisms for privacy-preserving mining.
Specifically, we demonstrate that (a) the prior techniques differ only
in their settings for the model parameters, and (b) through appropriate
choice of parameter settings, we can derive new perturbation techniques
that provide highly accurate mining results even under strict privacy
guarantees.  We also propose a novel perturbation mechanism wherein
the model parameters are themselves characterized as random variables,
and demonstrate that this feature provides significant improvements
in privacy at a very marginal cost in accuracy.

While our model is valid for random-perturbation-based privacy-preserving
mining in general, we specifically evaluate its utility here with regard
to frequent-itemset mining on a variety of real datasets. The experimental
results indicate that our mechanisms incur substantially lower identity
and support errors as compared to the prior techniques.

\end{abstract}

\jhcomment{In last line above, add info about privacy improvement?}

\Section{Introduction}

The knowledge models produced through data mining techniques are
only as good as the accuracy of their input data.  One source of data
inaccuracy is when users, due to privacy concerns, deliberately provide
wrong information.  This is especially common with regard to customers
asked to provide personal information on web forms to e-commerce
service providers.

The compulsion for doing so may be the (perhaps well-founded) worry that
the requested information may be misused by the service provider to harass
the customer.  As a case in point, consider a pharmaceutical company that
asks clients to disclose the diseases they have suffered from in order
to investigate the correlations in their occurrences -- for example,
``Adult females with malarial infections are also prone to contract
tuberculosis''. While the company may be acquiring the data solely for
genuine data mining purposes that would eventually reflect itself in
better service to the client, at the same time the client might worry
that if her medical records are either inadvertently or deliberately
disclosed, it may adversely affect her employment opportunities.

To encourage users to submit correct inputs, a variety of
privacy-preserving data mining techniques have been proposed in the last
few years~\cite{as00,condensation,gehrke02,mask,k-anony}.  The goal of these techniques is to keep
the raw local data private but, at the same time, support accurate reconstruction
of the global data mining models.  Most of the techniques are based
on a \emph{data perturbation} approach, wherein the user data is
distorted in a probabilistic manner that is disclosed to the eventual
miner.  For example, in the MASK technique~\cite{mask}, intended for
privacy-preserving association-rule mining on sparse boolean databases,
each $0$ or $1$ in the original user transaction vector is flipped with a
parametrized probability $1-p$.  \comment{Setting $p = 0.9$ was found to provide
a reasonable tradeoff between the conflicting goals of user privacy and
mining accuracy for the datasets considered in their evaluation.}

\jhcomment{include the list of references at end of first line.}

\SubSection{The FRAPP Framework}
The trend in the prior literature has been to propose \emph{specific}
perturbation techniques, which are then analyzed for their privacy
and accuracy properties.  We move on, in this paper, to presenting 
FRAPP (FRamework for Accuracy in Privacy-Preserving mining),
a generalized matrix-theoretic \emph{framework} that facilitates a
systematic approach to the design of random perturbation schemes for
privacy-preserving mining.  While various privacy metrics have been
discussed in the literature, FRAPP supports a particularly strong notion
of privacy, originally proposed in \cite{lim03}.  Specifically, it supports
a measure called ``amplification'', which guarantees strict limits on
privacy breaches of individual user information, \emph{independent of the 
distribution of the original data}.

FRAPP quantitatively characterizes the \emph{sources of error} in
random data perturbation and model reconstruction processes.  We first
demonstrate that the prior techniques differ only in their settings for
the FRAPP parameters. Further, and more importantly, we show that through
appropriate choice of parameter settings, new perturbation techniques can
be constructed that provide highly accurate mining results even under
strict privacy guarantees.  Efficient implementations
for these new perturbation techniques are also presented.

We investigate here, for the first time, the possibility of \emph{randomizing
the perturbation parameters themselves}. The motivation is that it could
lead to an increase in privacy levels since the exact parameter values
used by a specific client will not be known to the data miner.  This
scheme has the obvious downside of perhaps reducing the model reconstruction accuracy.
However, our investigation shows that the tradeoff is very attractive
in that the privacy increase is substantial whereas the accuracy
reduction is only marginal.  This opens up the possibility of using
FRAPP in a \emph{two-step} process: First, given a user-desired level of privacy,
identifying the deterministic values of the FRAPP parameters that
both guarantee this privacy and also maximize the accuracy; and then,
(optionally) randomizing these parameters to obtain even better privacy
guarantees at a minimal cost in accuracy.

The FRAPP model is valid for random-perturbation-based privacy-preserving
mining in general.  Here, we focus on its applications to
\emph{categorical databases}, where the domain of each attribute
is finite.  Note that boolean data is a special case of this class,
and further, that continuous-valued attributes can be converted into
categorical attributes by partitioning the domain of the attribute into
fixed length intervals.

To quantitatively evaluate FRAPP's utility, we specifically evaluate the
performance of our new perturbation mechanisms on the popular mining task
of finding frequent itemsets, the cornerstone of association rule mining.
Our evaluation on a variety of real datasets shows that both identity
and support errors are substantially lower than those incurred by the
prior privacy-preserving techniques.

\jhcomment{Want to add a line on privacy increase at end of above para?}


\comment{
Thus privacy concerns may influence users to provide spurious information.
Since the primary task in data mining is the development of models
about aggregated data, there was shown a potential of development of
accurate models for mining in \cite{as00} without access to precise info
in individual records ; thus preserving the client's data privacy. Since
then, there has been research on design of schemes which could randomly
perturb the user data at user end itself and at the same time allow the
miner to estimate the models about original data from the perturbed data.


Some data distortion schemes based on random perturbation were proposed
recently \cite{mask,gehrke02}.  In \cite{lim03}, the authors showed
that random perturbation approach is potentially vulnerable to privacy
breaches: based on the distribution of data, one may be able to learn with
high confidence that some of the randomized records satisfy a specified
property, even though privacy is preserved on average. They presented a
measure called "amplification" which guarantees limits on privacy breaches
without any knowledge of the distribution of the original data. We call
such privacy guarantees as \emph{strict} privacy guarantees.

In this paper we design random perturbation schemes to significantly
raise the mining accuracy bar under such \emph{strict} privacy
guarantees. Rather than choosing a perturbation method and then
analyzing the reconstruction error and privacy breaches for it, as is
done in most of the prior work, we develop a generalized theoretical
model to quantitatively characterize the sources of error in random
perturbation/reconstruction process under strict privacy guarantees. We
show that various techniques differ only in their choice of the parameters
of this model and derive  highly accurate privacy preserving mining
mechanisms through appropriate choice of the model parameters. We also
devise practical efficient perturbation algorithm for the proposed scheme.

}



\SubSection{Contributions}
In a nutshell, FRAPP provides a mathematical foundation for ``raising
both the accuracy and privacy bars in strict privacy-preserving mining''. 
Specifically, our main contributions are as follows:
\begin{itemize}
\item
FRAPP, a generalized matrix-theoretic framework for random perturbation and mining model reconstruction;
\item
Using FRAPP to derive new perturbation mechanisms for minimizing the model reconstruction error while ensuring strict privacy guarantees;
\item
Introducing the concept of randomization of perturbation parameters, and thereby deriving enhanced privacy;
\item
Efficient implementations of the perturbation techniques for the proposed mechanisms;
\item
Quantitatively demonstrating the utility of our schemes in the context of association rule mining.
\end{itemize}

\SubSection{Organization}

The remainder of this paper is organized as follows: 
The FRAPP framework for data perturbation and model reconstruction
is presented in Section \ref{sec:frapp}.  Appropriate choices
of FRAPP parameters for simultaneously guaranteeing strict data
privacy and providing high model accuracy are discussed in Section
\ref{sec:choiceA}.  The impact of randomizing the FRAPP parameters
is investigated in Section \ref{sec:randomA}.  Efficient schemes for
implementing the new perturbation mechanisms are described in Section
\ref{sec:perturbation_algo}.  In Section~\ref{sec:FI_mining}, we discuss
the application of our mechanisms to association rule mining.
Then, in Section~\ref{sec:experiments2},
the utility of FRAPP in the context of association rule mining is
quantitatively investigated.  Related work on privacy-preserving
mining is reviewed in Section~\ref{sec:related}.  Finally, in
Section~\ref{sec:conc}, we summarize the conclusions of our study
and outline future research avenues.



\Section{The FRAPP Framework}
\label{sec:frapp}
In this section, we describe the construction of the FRAPP framework,
and its quantification of privacy and accuracy measures.

\paragraph{Data Model.}
We assume that the original database $U$ consists of $N$ records, with
each record having $M$ categorical attributes.  
The domain of attribute $j$ is denoted by $S_{U}^j$,
resulting in the domain $S_U$ of a record in $U$ being given by
$\displaystyle S_U={\tiny \prod}_{j=1}^{M}S_U^j$. 
We map the domain $S_U$ to index set $ I_U=\{1,\ldots,|S_U|\} $, so that we can model 
the database as set of $N$ values from $I_U$. 
Thus, if we denote $i^{th}$ record of $U$ as $U_i$, we have

\begin{displaymath}
	U=\{U_i\}_{i=1}^{N}, \qquad U_i\in I_U
\end{displaymath}

\paragraph{Perturbation Model} 
We consider the privacy situation wherein the customers trust \emph{no
one except themselves}, that is, they wish to perturb their records at
their client site before the information is sent to the the miner,
or any intermediate party.  This means that  perturbation is done at the level
of \emph{individual} customer records $U_i$, without being influenced by the
contents of the other records in the database.

For this situation, there are two possibilities: a simple
\emph{independent column perturbation}, wherein the value of each
attribute in the record is perturbed independently of the rest, or
a more generalized \emph{dependent column perturbation}, where the
perturbation of each column may be affected by the perturbations of the
other columns in the record. Most of the prior perturbation techniques,
including \cite{gehrke02,lim03,mask}, fall into the independent column
perturbation category. The FRAPP framework, however, includes both kinds
of perturbation in its analysis.

\comment{
If we define \emph{granularity of perturbation} as the part of the
database which is distorted independent of the rest of the database ,
or in more intuitive terms, without seeing the rest of the database,
then the maximum granularity of perturbation for this privacy model is
a single record $U_i$.  For the column independent perturbation schemes
like those used in prior work \cite{mask,gehrke02,lim03}, the value of
each column for each record , ie. $U_{ij}, j\in\{1,....M\}$ is maximum
granularity of perturbation. The privacy model actually allows methods
where perturbation of the value of one column in the record can be
dependent on perturbation of other columns in the record. So we will
assume $U_i$ as the granularity of perturbation for generalized analysis.
}

Let the perturbed database be $V=\{V_1,\ldots,V_N\}$, with domain $S_V$, and corresponding index set $I_V$. 
For each original customer record $U_i=u, u \in I_U$, a  new perturbed
record $V_i=v, v \in I_V$ is randomly generated with probability $p(u \to v)$.
Let $A$ denote the matrix of these transition probabilities, with
$\displaystyle A_{vu} =  p(u \to v)$.
This random process maps to a Markov process, and the perturbation matrix $A$ 
should therefore satisfy the following properties \cite{bookGstrang}: 	
{\setlength\arraycolsep{1.0pt}
       \begin{eqnarray}
	\label{Aeq}
	\sum_{v\in I_V}{A_{vu}} & = & 1 \qquad\forall u\in I_U \nonumber\\
	A_{vu} & \geq  & 0 \qquad \forall u \in I_U,v \in I_V 
	\end{eqnarray}
}
\jhcomment{The exists symbol does not make sense to me.}

\noindent
Due to the constraints imposed by Equation \ref{Aeq}, the domain of
$A$ is not $\mathbf{R}^{|S_U| \times \mathbf |S_V|}$ but a subset of
it.  This domain is further restricted
by the choice of perturbation method.  For example, for the MASK
technique~\cite{mask} mentioned in the Introduction, all the entries of
matrix $A$ are decided by the choice of the single parameter $p$.

In this paper, we propose to explore the \emph{preferred choices} of $A$ to 
simultaneously achieve privacy guarantees and high accuracy, without 
restricting ourselves ab initio to a particular perturbation method.


\SubSection{Privacy Guarantees}
\label{sec:privacyguarantees}

The miner receives the perturbed database $V$ and attempts to reconstruct
the original probability distribution of database $U$ using this perturbed
data and the knowledge of the perturbation matrix $A$.

The \emph{prior probability} of a property of a customer's private
information is the likelihood of the property in the absence of any
knowledge about the customer's private information. On the other hand,
the \emph{posterior probability} is the likelihood of the property given
the perturbed information from the customer and the knowledge of the
prior probabilities through reconstruction from the perturbed database.
As discussed in \cite{lim03}, in order to preserve the privacy of some
property of a customer's private information, we desire that the posterior
probability of that property should not be much higher than the prior
probability of the property for the customer.  This is quantified by
saying that a perturbation method has privacy guarantees $(\rho_1,\rho_2)$
if, for any property $Q(U_i)$ with prior probability less than $\rho_1$,
the posterior probability of the property is guaranteed to be less than $\rho_2$.

For our formulation, we derive (using Definition 3 and Statement 1 from \cite{lim03})
the following condition on the perturbation matrix $A$ in order to support
$(\rho_1,\rho_2)$ privacy.
\begin{equation} \label{pconstraint}
\frac{A_{vu_1}}{A_{vu_2}} \leq \gamma \leq \frac{\rho_2(1-\rho_1)}{\rho_1(1-\rho_2)} \qquad u_1,u_2 \in I_U, \forall v\in I_V 
\end{equation}

\noindent
That is, the choice of perturbation matrix $A$ should follow the restriction that
the ratio of any two entries should not be more than $\gamma$.

\SubSection{Reconstruction Model}
\label{sec:model}

We now analyze how the distribution of the original database can be reconstructed
from the perturbed database.
As per the perturbation model, a client $C_i$ with data record $U_i=u, u\in I_U$
generates record $V_i=v, v\in I_V$ with probability $p[u \rightarrow v]$. 
This event of generation of $v$ can be viewed as a Bernoulli trial with success probability $p[u \rightarrow v]$. 
If we denote outcome of $i^{th}$ Bernoulli trial by random variable
$Y_v^i$, then the total number of successes $Y_v$ in $N$ trials is given by sum
of the $N$ Bernoulli random variables.  i.e.  
\begin{equation}
\label{sum_Yv}
Y_v=\sum_{i=1}^{N}{Y_v^i} 
\end{equation}
That is, the total number of records with value $v$ in the perturbed database will be 
given by the total number of successes $Y_v$.

Note that $Y_v$ is the sum of $N$ independent but non-identical Bernoulli
trials. The trials are non-identical because the probability of success in a
trial $i$ varies from another trial $j$ and actually depends on the values of
$U_i$ and $U_j$, respectively. The distribution of such a random variable $Y_v$
is known as the Poisson-Binomial distribution \cite{yhwang}.

Now, from Equation~\ref{sum_Yv}, the expectation of $Y_v$ is given by
\begin{equation} \label{expectationv0}
	E(Y_v)=\sum_{i=1}^{N}{E(Y_v^i)}=\sum_{i=1}^{N}{P(Y_v^i=1)}
\end{equation}
Let $X_u$ denote the number of records with value $u$ in the original database.
Since \mbox{$P(Y_v^i=1)=p[u \rightarrow v]={A_{vu}}$, for $U_i=u$}, we get
\begin{eqnarray}
\label{expectationv}
E(Y_v) =\sum_{u \in I_U}A_{vu}X_u
\end{eqnarray}
Let $X=[{X_1} X_2 \cdots {X_{|S_U|}}]^T$, $Y=[Y_1 Y_2 \cdots Y_{|S_V|}]^T$,
then from Equation~\ref{expectationv} we get
\begin{equation}
\label{estimation_det_eq}
E(Y)=AX 
\end{equation}
We estimate $X$ as $\widehat{X}$ given by the solution of following equation
\begin{equation}
\label{approximate_det_eq0}
	Y=A\widehat{X}
\end{equation}
which is an approximation to Equation~\ref{estimation_det_eq}.
This is a system of $|S_V|$ equations in $|S_U|$ unknowns.
For the system to be uniquely solvable, a necessary condition 
is that the space of the perturbed database is larger 
than or equal to the original database (i.e. $|S_V| \geq |S_U|$).
Further, if the inverse of matrix $A$ exists, then we can find the solution of above system of equations by
\begin{equation}
\label{approximate_det_eq}
\widehat{X}=A^{-1}Y
\end{equation}
That is, Equation~\ref{approximate_det_eq} gives the estimate of the
distribution of records in the original database, which is the objective
of the reconstruction exercise.


\SubSection{Estimation Error}
To analyze the error in the above estimation process, we use the following well-known theorem
from linear algebra~\cite{bookGstrang}:\\
\\
{\bf Theorem 1:}
For an equation of form $Ax=b$, the relative error in solution $x=A^{-1}b$ satisfies 
\[
\frac{\parallel \delta x \parallel}{\parallel x \parallel} \leq c \frac{\parallel \delta b \parallel}{\parallel b \parallel} 
\]
where $c$ is the \emph{condition number} of matrix $A$. For a positive
definite matrix, $c = \lambda_{max}/\lambda_{min}$, where $\lambda_{max}$
and $\lambda_{min}$ are the maximum and minimum eigen values of $n\times n$ matrix $A$.
Informally, the condition number is a measure of stability or sensitivity
of a matrix to numerical operations.  Matrices with condition numbers
near one are said to be \emph{well-conditioned}, whereas those with condition
numbers much greater than one (e.g. $10^5$ for a $5 * 5$ Hilbert matrix~\cite{bookGstrang})
are said to be \emph{ill-conditioned}.

\jhcomment{Which norm is being used in above theorem?}

From Equations \ref{estimation_det_eq}, \ref{approximate_det_eq} and the above theorem, we have
\begin{equation}
\frac{\parallel \widehat{X}-X \parallel}{\parallel X \parallel} \leq c\frac{\parallel Y-E(Y)\parallel}{\parallel E(Y) \parallel}
\end{equation} 
This inequality means that the error in estimation arises from two
sources: First, the sensitivity of the problem which is measured by
the condition number of matrix $A$; and, second, the deviation of $Y$
from its mean as measured by the  variance of $Y$.

As discussed above, $Y_v$ is a Poisson-Binomial distributed random
variable. Hence, using the expression for variance of a Poisson-Binomial
random variable~\cite{yhwang}, we can compute the variance of $Y_v$ to be

\begin{eqnarray}
\label{variancev0}
Var(Y_v) & = & A_vX(1-\frac{1}{N}A_vX) \nonumber\\
{} & {} &- \sum_{u\in I_U}(A_{vu}-\frac{1}{N}A_vX)^2{X_u}
\end{eqnarray}
which depends on the perturbation matrix $A$ and the distribution $X$ of records in the
original database.  Thus the effectiveness of the privacy preserving method
is \emph{critically dependent on the choice of matrix $A$}.


\Section{Choice of Perturbation Matrix} \label{sec:choiceA}

The various perturbation techniques proposed in the literature 
primarily differ in their choice for perturbation matrix $A$.  For example,
\begin{itemize}
\item
MASK \cite{mask} uses the matrix $A$ with
\begin{eqnarray}
\label{matrix_MASK}
	A_{vu}  =  p^k(1-p)^{M_b-k}
\end{eqnarray}
where $k$ is the number of attributes with matching values in perturbed value $v$
and original value $u$, $M_b$ is the number of boolean attributes when each categorical
attribute $j$ is converted into $\mid S_U^j\mid$ boolean attributes, and
$1-p$ is the value flipping probability.

\item
The \emph{cut-and-paste} randomization operator \cite{gehrke02}
employs a matrix $A$ with
\begin{small}
\begin{eqnarray}
\label{matrix_cut_paste}
	A_{vu} & = & \sum_{z=0}^{M} p_M[z] \sum_{q=max\{0,z+l_u-M,l_u+l_v-M_b\}}^{min\{z,l_u,l_v\}} \frac{{\choose{l_u}{q}}{\choose{M-l_u}{z-q}}}{\choose{M}{z}}\nonumber\\ 
	{} & {} & \qquad \cdot {\choose{M_b-l_u}{l_v-q}} \rho^{(l_v-q)}(1-\rho)^{(M_b-l_u-l_v+q)} \nonumber\\
\textrm{where}
\end{eqnarray}
\end{small}
\begin{small}
\begin{eqnarray}
p_M[z] & = & \sum_{w=0}^{min\{K,z\}}{\choose{M-w}{z-w}}\rho^{(z-w)}(1-\rho)^{(M-z)} \nonumber\\
{} & {} & \cdot \left\{ \begin{array}{ll} 1-M/(K+1) & \textrm{if } w=M \textrm{ and } w<K \\ 1/(K+1) & \textrm{o.w.} \end{array} \right. \nonumber
\end{eqnarray}
\end{small}
Here $l_u$ and $l_v$ are the number of $1^s$ in the original record $u$
and its corresponding perturbed record $v$, respectively, while $K$ and $\rho$ are 
operator parameters.
\end{itemize}

\jhcomment{Wouldn't it be better to swap i and j to make the notation easy?}
\jhcomment{Would be better to have $l_u$ and $l_v$ instead of l and l'.}

For enforcing strict privacy guarantees, the parameters for the above
methods are decided by the constraints on the values of perturbation matrix $A$
given in Equation \ref{pconstraint}.  It turns out that for practical
values of privacy requirements, the resulting matrix $A$ for these
schemes is extremely \emph{ill-conditioned} -- in fact, we found the condition
numbers in our experiments to be of the order of $10^5$ and $10^7$
for MASK and the Cut-and-Paste operator, respectively. 

Such ill-conditioned matrices make the reconstruction very sensitive
to the variance in the distribution of the perturbed database. Thus,
it is important to carefully choose the matrix $A$ such that it is
well-conditioned (i.e has a low condition number).  If we decide on a
distortion method apriori, as in the prior techniques, then there is
little room for making specific choices of perturbation matrix $A$.
Therefore, we take the opposite approach of first designing matrices
of the required type, and then devising perturbation methods that are
compatible with the chosen matrices.

\jhcomment{What is "generated distribution" referring to here?}

\jhcomment{Do you want to call the perturbation matrix P throughout instead of A?}

To choose the appropriate matrix, we start from the intuition that for
$\gamma=\infty$, the matrix choice would be the unity matrix, which
satisfies the constraints on matrix $A$ imposed by Equations \ref{Aeq}
and  \ref{pconstraint},  and has condition number 1.
Hence, for a given $\gamma$, we can choose the following matrix:
\begin{eqnarray} \label{matrixA}
	A_{ij}  =  \left\{ 
		\begin{array}{ll} 
		\gamma x,  & \textrm{if } i=j \nonumber\\
		x, & \textrm{o.w.}
		\end{array}
		\right.		
		\nonumber\\
	where \nonumber\\
	x=\frac{1}{\gamma + (|S_U|-1)}
\end{eqnarray}
This matrix will be of the form 
\begin{displaymath}
\begin{array}{cc}
	x &
	\left[ \begin{array}{cccc}
		\gamma & 1 & 1 & \ldots\\
		1 & \gamma & 1 & \ldots\\
		1 & 1 & \gamma & \ldots\\
		\vdots & \vdots & \vdots & \ddots
		\end{array}
	\right]
\end{array}
\end{displaymath}
It is easy to see that the above matrix, which incidentally is a symmetric
Toeplitz matrix~\cite{bookGstrang}, satisfies the conditions given by
Equations~\ref{Aeq} and \ref{pconstraint}.  Further, its condition number
can be computed to be $1 + \frac{\mid S_U \mid}{\gamma - 1}$, \comment{which will
clearly be small for most practical mining environments.}
For ease of exposition, we will hereafter refer to this matrix informally 
as the ``gamma-diagonal matrix''.  

At this point, an obvious question is whether it is possible to design
matrices that have even lower condition number than the gamma-diagonal matrix.
In the remainder of this section, we prove that within the constraints of
our problem, the gamma-diagonal
matrix has the \emph{lowest} possible condition number, that is, it is
an \emph{optimal choice} (albeit non-unique).

\paragraph{Proof.}
To prove this, we will first derive the expression for minimum condition
number for such matrices and the conditions under which that condition
number is achieved. Then we show that our gamma-diagonal matrix
satisfies these conditions, and has minimum condition number.

For a symmetric positive definite matrix, the condition number is given by 
\begin{eqnarray}
c & = & \frac{\lambda_{max}}{\lambda_{min}}
\end{eqnarray}
where $\lambda_{max}$ and $\lambda_{min}$ are the maximum and minimum eigenvalues of the matrix.
As the matrix $A$ is a Markov matrix (refer Equation~\ref{Aeq}), the following
theorem for eigenvalues of a matrix can be used \\
\\
{\bf Theorem 2} \cite{bookGstrang} {\em For an $n\times n$ Markov matrix,
\begin{itemize} 
\item{$1$ is an eigenvalue}
\item{the other eigenvalues satisfy $\mid \lambda_i \mid \le 1$} 
\end{itemize}}

\noindent
{\bf Theorem 3} \cite{bookGstrang} {\em The sum of $n$ eigenvalues equals the sum of $n$ diagonal entries:
\begin{equation*}
\lambda_1+\cdots+\lambda_n=A_{11}+\cdots+A_{nn}
\end{equation*}}
\\
Using Theorem 2 we get,
\begin{eqnarray}
\label{lambda_results}
\lambda_{max} & = & 1 \nonumber
\end{eqnarray}
\\
As the least eigenvalue $\lambda_{min}$ will always be less than or equal to average of the eigenvalues other than $\lambda_{max}$, we get,
\begin{eqnarray}
\label{lmin_bound}
\lambda_{min} & \le & \frac{1}{n-1}{\sum_{i=2}^{n}\lambda_i} \nonumber
\end{eqnarray}
where $\lambda_1=\lambda_{max}$
Using Theorem 3,
\begin{eqnarray}
\lambda_{min} & \le &\frac{1}{n-1}\Bigg(\sum_{i=1}^{n}A_{ii}-1\Bigg),
\end{eqnarray}

Hence, condition number, 
\begin{equation} 
\label{c_bound1}
c = \frac{1}{\lambda_{min}} \ge \frac{n-1}{\sum_{i=1}^{n}A_{ii}-1}
\end{equation}
Now, due to privacy constraints on $A$ given by Equation~\ref{pconstraint},
\[A_{ii} \le \gamma A_{ij} \textrm{ for any } j\ne i,\] i.e.,
\begin{displaymath}
 \begin{array}{ccc}
  	A_{ii} & \le & \gamma A_{i1}\\
  	A_{ii} & \le & \gamma A_{i2}\\
  	{} & \vdots & {}
  	\end{array}
\end{displaymath}
Summing above,
\begin{eqnarray}
(n-1)A_{ii} & \le & \gamma \sum_{j \ne i}A_{ij} \nonumber\\
& = & \gamma(1-A_{ii})\nonumber
\end{eqnarray}
where the last step is due to the condition on $A$ given by Equation~\ref{Aeq}.
Solving for $A_{ii}$, we get,
\begin{eqnarray}
A_{ii} & \le & \frac{\gamma}{\gamma+n-1}
\end{eqnarray}
Using above inequality in Equation~\ref{c_bound1}, we get
\begin{equation}
\label{c_bound2}
c \ge \frac{n-1}{\frac{n\gamma}{\gamma+n-1}-1}=\frac{\gamma+n-1}{\gamma-1}
\end{equation}

Hence minimum condition number for the symmetric perturbation matrices under
privacy constraints represented by $\gamma$ is $\frac{\gamma+n-1}{\gamma-1}$.
This condition number is achieved when \mbox{$A_{ii}=\frac{\gamma}{\gamma+n-1}$}. 

The diagonal values of gamma-diagonal matrix given by
Equation~\ref{Aeq} is $\frac{\gamma}{\gamma+n-1}$. Thus it is
\emph{minimum condition number} symmetric perturbation matrix, with
condition number $\frac{\gamma+\mid S_U\mid-1}{\gamma-1}$.

\Section{Randomizing the Perturbation Matrix}
\label{sec:randomA}

The estimation model in the previous section implicitly assumed the
perturbation matrix $A$ to be \emph{deterministic}.  However, it appears
intuitive that if the perturbation matrix parameters are themselves
\emph{randomized}, so that each client uses a perturbation matrix that
is not specifically known to the miner, the privacy of the client will
be further increased.  Of course, it may also happen that the 
reconstruction accuracy may suffer in this process.

In this section, we explore this tradeoff. Instead of deterministic matrix $A$, the perturbation matrix here is matrix
$\tilde{A}$ of random variables, where each entry $\tilde{A}_{vu}$ is a random vaiable with $E(\tilde{A}_{vu})=A_{vu}$.  The values taken by the random variables for a client $C_i$ provide the specific values for
his/her perturbation matrix.

\SubSection{Privacy Guarantees}
\label{sec:privacy_rm}
Let $Q(U_i)$ be a property of client $C_i$'s private information, and
let record $U_i=u$ be perturbed to $V_i=v$. Denote the prior probability
of $Q(U_i)$ by $P(Q(U_i))$.  On seeing the perturbed data, the posterior
probability of the property is calculated to be:
\begin{small}
\begin{eqnarray}
P(Q(U_i)|V_i=v) & = & \sum_{Q(u)}P_{U_i|V_i}(u|v) \nonumber\\
		& = & \sum_{Q(u)}\frac{P_{U_i}(u)P_{V_i|U_i}(v|u)}{P_{V_i}(v)} \nonumber 
\end{eqnarray}
\end{small}

\noindent
When we use a fixed perturbation matrix $A$ for all clients $i$, then \mbox{$P_{V_i/U_i}(v/u)=A_{vu}, \forall i$}.
Hence 
\begin{small}
{\setlength\arraycolsep{1.0pt}
\begin{eqnarray}
P(Q(U_i)|V_i=v) & = & \frac{\sum_{Q(u)}P_{U_i}(u)A_{vu}}{\sum_{Q(u)}P_{U_i}(u)A_{vu} + \sum_{\neg Q(u)}P_{U_i}(u)A_{vu}} \nonumber 
\end{eqnarray}}
\end{small}
\noindent
As discussed in \cite{lim03}, the data distribution $P_{U_i}$ in the worst case can be such that
$ P(U_i=u)>0$  only if 
\begin{small}
\mbox{$\{u \in I_U | Q(u) \textrm{ and } A_{vu}=maxp\}$}\\
or \\
\mbox{$\{u \in I_U | \neg Q(u) \textrm{ and } A_{vu}=minp\}$},\\
\end{small}
so that
\begin{small}
\begin{eqnarray}
P(Q(U_i)/V_i=v) = \frac{ P(Q(u)) \cdot maxp}{ P(Q(u)) \cdot maxp + P(\neg Q(u)) \cdot minp } \nonumber 
\end{eqnarray}
\end{small}
where \mbox{\small$ maxp=\max_{Q(u')}A_{vu'}$} and \mbox{\small $minp=\min_{\neg Q(u')}A_{vu'}$}.
Since the distribution $P_U$ is known through reconstruction to the
miner, and matrix $A$ is fixed, the above posterior probability can be
determined by the miner.  For example, if \mbox{$P(Q(u))=5\%, \gamma=19$}, the posterior probability can be computed to be $50\%$
for perturbation with the gamma-diagonal matrix.


But, in the randomized matrix case where $P_{V_i/U_i}(v/u)$ is a realization of random variable $\tilde{A}$, only its distribution and not the exact value for a given $i$ is known to the miner. 
Thus determinations like the above cannot be made by the miner for a 
given record $U_i$.  For example, suppose we choose matrix $A$ such that
\begin{eqnarray}
\label{matrixA_bar} 
	A_{uv}  =  \left\{ 
		\begin{array}{ll} 
		\gamma x  + r, & \textrm{if } u=v \nonumber\\
		x - \frac{r}{\mid S_U \mid -1}, & \textrm{o.w.}
		\end{array}
		\right.		
\end{eqnarray}
where $x = \frac{1}{\gamma + (|S_U|-1)}$ and $r$ is a random variable uniformly distributed between $[-\alpha,\alpha]$.
Thus, the worst case posterior probability for a record $U_i$ is now a function of the value of $r$, and is given by
\begin{small}
\begin{displaymath}
\rho_2(r) = \frac{ P(Q(u)) \cdot \gamma x + r}{ P(Q(u)) \cdot (\gamma x + r) +  P(\neg Q(u))( x -
\frac{r}{\mid S_U \mid -1} )}
\end{displaymath}
\end{small}
Therefore, only the posterior probability \emph{range}, i.e.
\mbox{$[\rho_2^-,\rho_2^+]=[\rho_2(-\alpha),\rho_2(+\alpha)]$}, and the
distribution over the range, can be determined by the miner.  For example, for
the situation \mbox{$P(Q(u))=5\%, \gamma=19, \alpha=\gamma x/2$}, he can only
say that the posterior probability lies in the range [$33\%,60\%$] with its
probability of being greater than 50\% ($\rho_2$ corresponding to $r=0$) equal
to its probability of being less than 50\%.

\jhcomment{How did you arrive at 50 \%? Is that statement necessary?}

\SubSection{Reconstruction Model}
The reconstruction model for the deterministic perturbation matrix $A$
was discussed in Section~\ref{sec:model}.  We now describe the 
changes to this analysis for the randomized perturbation matrix $\tilde{A}$.
\\
The probability of success for Bernoulli variable $Y_v^i$ is now modified to
\begin{displaymath}
	P(Y_v^i=1)=\tilde{A}_{vu}^i, \textrm{  for }U_i=u
\end{displaymath}
where $\tilde{A}_{vu}^i$ denotes the $i^{th}$ realization of random variable $\tilde{A}_{vu}$.

Thus, from Equation~\ref{expectationv0},
\begin{small}
\begin{eqnarray}
\label{expectationv_rm}
	E(Y_v) & = & \sum_{i=1}^{N}{P(Y_v^i=1)} \nonumber\\
	{} & = & \sum_{u\in I_U}{\sum_{\{i|U_i=u\}}{{\tilde{A}}_{vu}^i}} \nonumber\\
	{} & = & \sum_{u\in I_U}{\overline{A}_{vu} {X_u}} \\
	\Rightarrow E(Y) & = & \overline{A}_{vu} X
\end{eqnarray}
\end{small}
\noindent
where $\overline{A}_{vu}  = \frac{1}{{X_u}}\sum_{\{i|U_i=u\}}{\tilde{A}_{vu}^i}$
is the average of the values taken by $\tilde{A}_{vu}$ for the clients whose original data record had value $u$. 

$\tilde{A}_{vu}$ is a random variable with expectation \mbox{$E(\tilde{A}_{vu})=A_{vu}$}, it can be easily seen that,
\begin{equation}
\label{expectationA}
	E(\overline{A}_{vu})=A_{vu}
\end{equation}
Hence, from Equation~\ref{expectationv_rm}, we get

\comment{
\begin{equation}
	E(E(Y_v))=\sum_{u\in I_U}{{A}_{vu} \times {X_u}}	
\end{equation}

Following the same terminology as in Section \ref{sec:model}, the above equation implies that:
\begin{equation}
	E(Y)=\overline{A}X 
\end{equation}
}

\begin{equation}
\label{estimation_eq}
	E(E(Y))=AX 
\end{equation}
We estimate $X$ as $\widehat{X}$ given by the solution of following equation
\begin{equation}
	Y=A\widehat{X}
\end{equation}
which is an approximation to Equation~\ref{estimation_eq}.
From \mbox{\emph{Theorem 1}} in Section \ref{sec:model}, the error in estimation is bounded by:
\begin{small}
\begin{equation}
\frac{\parallel \widehat{X}-{X} \parallel}{\parallel X \parallel} \leq c\frac{\parallel Y-E(E(Y))\parallel}{\parallel E(E(Y)) \parallel}
\end{equation} 
\end{small}
where $\boldmath{c}$ is the condition number of perturbation matrix $A$.

We now compare these bounds with the corresponding bounds of the
deterministic case.  Firstly, note that, due to the use of the randomized
matrix, there is a \emph{double expectation} for $Y$ on the RHS of the
inequality, as opposed to the single expectation in the deterministic
case. Secondly, only the numerator is different between the two cases
since $E(E(Y))=AX$.  Now, we have
\noindent
\begin{small}
\begin{eqnarray}
& {} & {} \parallel Y-E(E(Y)) \parallel \nonumber\\
{} & = & \parallel(Y-E(Y)) + (E(Y)-E(E(Y))) \parallel \nonumber\\
{} & \leq & \parallel Y-E(Y)\parallel + \parallel E(Y)-E(E(Y)) \parallel \nonumber
\end{eqnarray}
\end{small}

\noindent
Here $\parallel Y-E(Y)\parallel$ is given by the variance of random variable $Y$.
Since $Y_v$, as discussed before, is Poisson-binomial distributed, its
variance is given by \cite{yhwang}
\comment{
\begin{eqnarray}
\label{expectationv_rm0}
E(Y_v) & = & \sum_i{p_v^i}\\
\textrm{and variance} & & \nonumber\\
\end{eqnarray}
}
\begin{small}
\begin{eqnarray}
\label{variancev0_rm}
Var(Y_v) & =  & N \overline{p}_v - \sum_i{(p_v^i)^2}
\end{eqnarray}
\end{small}
where \mbox{$\overline{p}_v = \frac{1}{N}\sum_i{p_v^i}$} and $p_v^i = P(Y_v^i=1)$.

It is easily seen (by elementary calculus or induction) that among all
combinations $\{p_v^i\}$ such that $\sum_i{p_v^i}=n\overline{p}_v$,
the sum $\sum_i{(p_v^i)^2}$ assumes its minimum value when all $p_v^i$
are equal. It follows that, if the average probability of success
$\overline{p}_v$ is kept constant, $Var(Y_v)$ assumes its maximum
value when $p_v^1 = \cdots = p_v^N$. In other words, the variability of
$p_v^i$, or \emph{its lack of uniformity, decreases the magnitude of chance
fluctuations}, as measured by its variance \cite{bookfeller}. On using random
matrix $\tilde{A}$ instead of deterministic $A$ we increase the variability
of $p_v^i$ (now $p_v^i$ assumes variable values for all $i$), hence
decreasing the fluctuation of $Y_v$ from its expectation, as measured
by its variance.

Hence, \mbox{\small $\parallel Y-E(Y)\parallel$} is likely to be decreased
as compared to the deterministic case, thereby reducing the error bound.
\comment{ This is because of
expected decrease in variance of $Y$ due to increase in variability of
probabilities of non-identical Bernoulli trials, as observed earlier in this
section. } On the other hand, the positive value \mbox{\small $\parallel E(Y)-E(E(Y))
\parallel=\parallel (\overline{A}-A)X \parallel$}, which depends upon the
variance of the random variables in $\tilde{A}$, was $0$ in the deterministic 
case. Thus, the error bound is increased by this term.  

So, we have a classic tradeoff situation here, and as shown later in our
experiments of Section~\ref{sec:experiments2}, the tradeoff turns out very
much in our favour with the two opposing terms almost canceling each other
out, making the error only marginally worse than the deterministic case.

\jhcomment{need to do better presentation of equations in this section.}
\jhcomment{So, what are the implications? Why are you only concerned
about numerator?}
\shcomment{add some intuition of behavior of variance}

\Section{Implementation of Perturbation Algorithm}
\label{sec:perturbation_algo}
To implement the perturbation process discussed in the previous sections,
we effectively need to generate for each \mbox{$U_i=u$}, a discrete
distribution with PMF \mbox{$P(v)=A_{vu}$} and CDF $F(v)=\sum_{i \leq
v}A_{iu}$, defined over $v=1,\ldots,\mid S_V\mid$. 

A straightforward algorithm
for generating the perturbed record $v$ from the original record $u$
is the following

\begin{enumerate}
\item Generate $r \sim {\cal U}(0,1)$ 
\item Repeat for $v=1,\ldots,\mid S_V \mid$ 
	\begin{description}
	\item if $F(v-1) < r \leq F(v)$ \hspace*{0.2in} 
	\item return $V_i=v$
	\end{description}
\end{enumerate}
where ${\cal U}(0,1)$ denotes uniform distribtion over range $[0,1]$
\jhcomment{need to format above pseudocode correctly.}

This algorithm, whose complexity is proportional to the \emph{product} of
the cardinalities of the attribute domains, will require $\mid S_V \mid/2$ iterations 
on average which can turn out to be very large. For example, with $31$ attributes,
each with two categories, this amounts to $2^{30}$ iterations for each
customer!  We therefore present below an alternative algorithm whose complexity
is proportional to the \emph{sum} of the cardinality of the attribute domains. 


Given that we want to perturb the record \mbox{$U_i=u$}, we can write 
{\setlength\arraycolsep{1.0 pt}
\begin{displaymath}
\begin{array}{l}
P(V_i;U_i=u)\\
= P(V_{i1},\ldots,V_{iM};u) \\
= P(V_{i1};u) \cdot P(V_{i2}|V_{i1};u) \cdots P(V_{iM}|V_{i1},\ldots,V_{i(M-1)};u)
\end{array}
\end{displaymath}
}
\comment{
\\
$P(V_i;U_i=u) $\\
$= P(V_{i1},\ldots,V_{iM};u)$ \\
$= P(V_{i1};u) \cdot P(V_{i2}|V_{i1};u) \cdots P(V_{iM}|V_{i1},\ldots,V_{i(M-1)};u)$
}
For the perturbation matrix $A$, we get the following expressions for
the above probabilities:
\begin{eqnarray}
P(V_{i1}=a;u) 		& = & \sum_{\{v|v(1)=a\}}{A_{vu}} \nonumber \\
P(V_{i2}=b|V_{i1}=a;u) 	& = & \frac{P(V_{i2}=b,V_{i1}=a;u)}{P(V_{i1}=a);u}\nonumber\\
			& = & \frac{\sum_{\{v|v(1)=a \& v(2)=b\}}{A_{vu}}}{P(V_{i1}=a;u)} \nonumber\\
& & \ldots \textrm{and so on } \nonumber
\end{eqnarray}
where $v(i)$ denotes value of $i^{th}$ column for record value =$v$.


For the gamma-diagonal matrix $A$, and using $n_j$ to represent
$\prod_{k=1}^{j}{\mid S_U^k\mid}$, we get the following  expressions for
these probabilities after some simple algebraic calculations:
\begin{eqnarray}
P(V_{i1}=b;U_{i1}=b) & = & (\gamma+\frac{n_M}{n_1}-1)x \nonumber\\
P(V_{i1}=b;U_{i1}\neq b) & = & \frac{n_M}{n_1}x \nonumber
\end{eqnarray}
Then, for the $j^{th}$ attribute 
\begin{eqnarray}
\label{perturbalgo}
P(V_{ij}/V_{i1},\ldots,V_{i(j-1)};U_{i}) \qquad \qquad \qquad \qquad \qquad & & \nonumber\\
= \left\{ 
 \begin{array}{ll}
	\frac{(\gamma+\frac{n_M}{n_j}-1)x}{\prod_{k=1}^{j-1}p_{k}} ,&\qquad \textrm{if }\forall k \le j,  V_{ik}=U_{ik} \\
	\\
	\frac{(\frac{n_M}{n_j})x}{\prod_{k=1}^{j-1}p_{k}} ,&\qquad \textrm{o.w.} \\
\end{array}
\right. \nonumber\\
\end{eqnarray}
where $p_{k}$ is the probability that $V_{ik}$ takes value $a$, 
given that $a$ is the outcome of the random process performed for $k^{th}$ attribute. i.e.
\[p_{k}=P(V_{ik}=a/V_{i1},\ldots,V_{i(k-1)};U_{i})\]

\jhcomment{Equation formatting is messed up.}

Therefore, to achieve the desired random perturbation for a value in
column $j$, we use as input both its original value and the perturbed value of
the previous column $j-1$, and generate the perturbed value as per the
discrete distribution given in Equation \ref{perturbalgo}.
Note that is an example of \emph{dependent column perturbation},
in contrast to the independent column perturbation used in most
of the prior techniques.

To assess the complexity, it is easy to see that the average number
of iterations for the $j^{th}$ discrete distribution will be ${\small
|S_U^j|/2}$, and hence the average number of iterations for generating a
perturbed record will be ${\small \sum_{j}|S_U^j|/2}$ (this value
turns out to be exactly $M$ for a boolean database).

\Section{Application to Association Rule Mining}
\label{sec:FI_mining}
To illustrate the utility of the FRAPP framework, we demonstrate in
this section how it can be used for enhancing privacy-preserving mining of
\emph{association rules}, a popular mining model that identifies interesting
correlations between database attributes~\cite{ais93,quanassoc96}.

\comment{
Let us treat a record in the database as a set of \mbox{(attribute,
value)} pairs. Let $ {\cal I}=\{1,2,\ldots,M\}$ denote the set
of categorical attributes in $U$. Let ${\cal I}_a$ denote the set
\mbox{${\cal I}_a=\{<i,a_i>|i \in {\cal I}, a_i \in s_{U}^i\}$}. Thus
a pair $<i,a_i>$ in ${\cal I}_a$ denotes attribute $i$, with
associated value $a_i$. For any ${\cal X} \subseteq { \cal I}_a $, let
$attributes({\cal X})$ denote the set $\{i|<i,a_i> \in {\cal X}\}$. ${\cal
X}$ is called an itemset over attributes $attributes({\cal X})$.

Thus each record in the database can be represented as set of $M$
\mbox{(attribute,value)} pairs from $I_a$. Let \mbox{$f_R : I_U
\rightarrow {\cal P}(I_a)$} \footnote{Set ${\cal P}(S)$ denotes power
set of set $S$} is a function which uniquely maps value $u \in I_U$ of a
record in dataset $U$ (Section \ref{sec:datamodel}) to the corresponding
set $R \subset {\cal I}_a$. We say that a record $U_i=u$ supports itemset
${\cal X \subseteq I}_a, \textrm{if } {\cal X}\subseteq f_R(u)$.

An association rule is a (statistical) implication of the form $ \cal
X \Longrightarrow Y$, where $ {\cal X, \cal Y \subset \cal I}_a $ and
$attributes({\cal X}) \cap attributes({\cal Y}) = \phi$. A rule $ \cal
X  \Longrightarrow Y $ is said to have a \emph{support} (or frequency)
factor \emph{s} iff at least ${s}\% $ of the records in $U$ support
$\cal X \cup Y$.  A rule $\cal X  \Longrightarrow Y$ is satisfied in the
database $U$ with a {\em confidence} factor \emph{c} iff at least $ c\%
$ of the records in $ U $ that support $ \cal X $ also support $ \cal
Y $.  Both support and confidence are fractions in the interval [0,1].
The support is a measure of statistical significance, whereas confidence
is a measure of the strength of the rule.

A rule is said to be ``interesting'' if its support and confidence are
greater than user-defined thresholds ${sup}_{min}$ and ${con}_{min}$,
respectively, and the objective of the mining process is to find all such
interesting rules.  It has been shown in \cite{ais93} that achieving
this goal is effectively equivalent to generating all subsets $\cal X$
of ${\cal I}_a $ that have support greater than ${sup}_{min}$ -- these
subsets are called \emph{frequent} itemsets.  Therefore, the mining
objective is, in essence, to efficiently discover all frequent itemsets
that are present in the database.

\SubSection{Mining from the Distorted Database}

Having established the equations for prior distribution estimation
from the perturbed database in Section \ref{sec:model}, we move on to
present the technique for using the distribution estimation for the
frequent itemset mining objectives mentioned in previous section. 
}

The core of the association rule mining is to identify 
``frequent itemsets'', that is, all those itemsets whose support (i.e. frequency)
in the database is in excess of a user-specified threshold.
Equation \ref{approximate_det_eq} can be directly used to estimate the support of
itemsets containing all $M$ categorical attributes. However, in order
to incorporate the reconstruction procedure into bottom-up association rule
mining algorithms such as \emph{Apriori} \cite{as94}, we need to also be
able to estimate the supports of itemsets consisting of only a \emph{subset} of attributes.

Let $C$ denotes the set of all attributes in the database, and $C_s$ be a subset of
attributes. Each of the attributes $j\in C_s$ can assume one of the $|S_U^j|$
values. Thus, the number of itemsets over attributes in $C_s$ is given by
$n_{C_s}=\prod_{j\in{C_s}}|S_U^j|$. Let ${\cal L,H}$ denote itemsets over this
subset of attributes.

We say that record supports an
itemset $\cal L$ over $C_s$ if the entries in the record for the attributes
$j\in C_s$ are same as in $\cal L$.


Let \emph{support} of an itemset ${\cal L}$
in original and distorted database be denoted by $sup^U_{{\cal L}}$ and
$sup^V_{{\cal L}}$, respectively. 
Then,
\begin{eqnarray}
sup^V_{{\cal L}} & = & \frac{1}{N}\sum_{v \textrm{\tiny{ supports }} {\cal L}}Y_v \nonumber
\end{eqnarray}
where $Y_v$ denotes the number of records in $V$ with value $v$ (refer Section~\ref{sec:model}).
From Equation~\ref{approximate_det_eq0}, we know
\begin{eqnarray}
Y_v & = & \sum_{u\in I_U}A_{vu}\widehat{X}_u 
\end{eqnarray}
Hence,
\begin{eqnarray}
sup^V_{{\cal L}} & = & \frac{1}{N}\sum_{v \textrm{\tiny{ supports }} {\cal L}} \sum_{u} A_{vu}\widehat{X}_u  \nonumber\\
& = & \frac{1}{N} \sum_{u} \widehat{X}_u \sum_{v \textrm{\tiny{ supports }} {\cal L}}  A_{vu} \nonumber \\
& = & \frac{1}{N} \sum_{{\cal H}} \sum_{u \textrm{\tiny{ supports }} {\cal H}} \widehat{X}_u \sum_{v \textrm{\tiny{ supports }} {\cal L}}  A_{vu} \nonumber 
\end{eqnarray}
If for all $u$ which support a given itemset ${\cal H}$, $\sum_{v \textrm{\tiny{ supports }} {\cal L}}  A_{vu}={\cal A}_{\cal HL}$, i.e. it is equal for all $u$ which support a given itemset, then the above equation can be written as:
\begin{eqnarray}
sup^V_{{\cal L}} & = & \frac{1}{N} \sum_{{\cal H}} {\cal A}_{{\cal HL}} \sum_{u \textrm{\tiny{ supports }} {\cal H}} \widehat{X}_u  \nonumber \\
 {} & = & \sum_{{\cal H}} {\cal A}_{{\cal HL}} \qquad \widehat{sup^U}_{{\cal H}} \nonumber
\end{eqnarray}

Now we find the matrix ${\cal A}$ for our gamma-diagonal matrix. 
Through some simple algebra, we get following matrix $\cal A$
corresponding to itemsets over subset $C_s$,
\jhcomment{above is unreadable.}
Hence, 

\begin{eqnarray}
\label{matrixAcal}
	{\cal A}_{\cal H L}  =  \left\{ 
		\begin{array}{ll} 
		\gamma x + (\frac{n_{C}}{n_{C_s}}-1 )x, & \textrm{if } {\cal H=L} \nonumber\\
		\frac{n_{C}}{n_{C_s}} x, & \textrm{o.w.}
		\end{array}
		\right.		
		\nonumber\\
\end{eqnarray}
Using the above $n_{C_s} \times n_{C_s}$ matrix we can estimate support
of itemsets over any subset $C_s$ of attributes. Thus our scheme can
be implemented on popular bottom-up association rule mining algorithms.


\Section{Performance Analysis}
\label{sec:experiments2}
We move on, in this section, to quantify the utility of the FRAPP framework
with respect to the privacy and accuracy levels that it can provide for
mining frequent itemsets.

\begin{table*}[!htb]
\small
\caption{ \small{CENSUS Dataset}
\label{tab:categories_adult}}
\vspace{0.25cm}
\begin{center}
\begin{tabular}{|c|c|}
\hline
\emph{Attribute} & \emph{Categories} \\
\hline
age & $(15-35],(35-55],(55-75],>75$ \\
\hline
fnlwgt & $(0-1e5],(1e5-2e5],(1e5-3e5],(3e5-4e5],>4e5$ \\
\hline
hours-per-week & $(0-20],(20-40],(40-60],(60-80],>80$ \\
\hline
race & White, Asian-Pac-Islander, Amer-Indian-Eskimo, Other, Black \\
\hline
sex & Female, Male \\
\hline	
native-country & United-States, Other \\
\hline
 \end{tabular}
\end{center}
\end{table*}

\begin{table*}[!htb]
\small
\begin{center}
\caption{\small{HEALTH Dataset}
\label{tab:categories_health}}
\vspace{0.25cm}
\begin{tabular}{|c|c|}
\hline
\emph{Attribute} & \emph{Categories} \\
\hline
AGE (Age) & $[0-20),[20-40),[40-60),[60-80), \ge 80)$ \\
\hline
BDDAY12 (Bed days in past 12 months) & $[0-7),[7-15),[15-30),[30-60),\ge 60$ \\
\hline
DV12 (Doctor visits in past 12 months)	& $[0-7),[7-15),[15-30),[30-60),\ge 60$ \\ 
\hline
PHONE (Has Telephone)	& Yes,phone number given; Yes, no phone number given; No \\
\hline
SEX (Sex) & Male ; Female \\
\hline
INCFAM20 (Family Income) &  Less than \$20,000; \$20,000 or more \\
\hline
HEALTH (Health status) & Excellent; Very Good; Good; Fair; Poor \\
\hline
\end{tabular}
\end{center}
\end{table*}

\comment{
Most of the real world data used for mining consists of many categorical or
quantitative attributes.  But privacy constraints are not there on all of the
attributes in the datasets. Hence in our experiments we perform privacy
preserving mining on datasets where the number of attributes is not very high
but each attribute can take values from multiple categories, the privacy
requirement on the attributes is strict, and high accuracy for even long length
patterns is desired.
}

\shcomment{\small{Shall I include the discussion that what will be the reconstruction method 
when some columns are perturbed and some are not?}}

\paragraph{Datasets.}
We use the following real world datasets in our experiments:
\begin{description} 
\item[CENSUS]:~This dataset contains census information for 
approximately 50,000 adult American citizens.  It is available from the UCI
repository~\cite{uciml}, and is a popular benchmark in data mining
studies.  It is also representative of a database where there are fields
that users may prefer to keep private -- for example, the ``race''
and ``sex'' attributes.  We use three continuous ({\tt age, fnlwgt,
hours-per-week}) and three nominal attributes ({\tt native-country, sex,
race}) from the census database in our experiments. The continuous
attributes are partitioned into (five) equiwidth intervals to convert them into
categorical attributes. The categories used for each attribute are listed
in Table \ref{tab:categories_adult}.

\item[HEALTH]:~This dataset captures health information for over 100,000
patients collected by the US government \cite{UScensus}. We selected 3
continuous and 4 nominal attributes from the dataset for our experiments.
The continuous attributes were partitioned into equi-width intervals to convert
them into categorical attributes.  The attributes and their categories
are listed in Table \ref{tab:categories_health}.
\end{description}

We evaluated the association rule mining accuracy of our schemes
on the above datasets for $sup_{min}=2\%$. \mbox{Table \ref{tab:fitems}}
gives the number of frequent itemsets
in the datasets for $sup_{min}=2\%$.  

\begin{table}[!ht]
\small
\caption{Frequent Itemsets for $sup_{min}=0.02$
\label{tab:fitems}}
\begin{center}
\begin{tabular}{|l|c|c|c|c|c|c|c|}
\hline
& \multicolumn{7}{c|}{Itemset Length} \\
& 1 & 2 & 3 & 4 & 5 & 6 & 7\\
\hline
CENSUS & 19 & 102 & 203 & 165 & 64 & 10 & --\\
HEALTH & 23 & 123 & 292 & 361 & 250 & 86 & 12\\
\hline
\end{tabular}
\end{center}
\end{table}

\comment{
\begin{table}[!t]
\begin{center}
\caption{\small{HEALTH Dataset: Frequent itemsets for $sup_{min}=0.02$}
\label{tab:fitems_health}}
\begin{tabular}{|c|c|c|c|c|c|c|c|}
\hline
itemset length & 1 & 2 & 3 & 4 & 5 & 6 & 7\\
\hline
\hline
\end{tabular}
\end{center}
\end{table}
}

\paragraph{Privacy Metric.}
The $(\rho_1,\rho_2)$ strict privacy measure from \cite{lim03} is used
as the privacy metric.  While we experimented with a variety of privacy
settings, due to space limitations, we present results here for a sample
$(\rho_1,\rho_2)=(5\%,50\%)$, which was also used in \cite{lim03}. This
privacy value results in $\gamma=19$.

\paragraph{Accuracy Metrics.}
We evaluate two kinds of mining errors, Support Error and Identity Error,
in our experiments: 
\begin{description}
\item[Support Error ($\rho$)]
This metric reflects the (percentage) average relative error in the
reconstructed support values for those itemsets that are correctly
identified to be frequent. Denoting the number of frequent itemsets by $|F|$, 
the reconstructed support by
$\widehat{sup}$ and the actual support by $sup$, the support error
is computed over all frequent itemsets as
\[
{\rho}  = \frac{1}{\mid F \mid}  \Sigma_{f\in F} 
\frac{\mid \widehat{sup}_f - sup_f \mid}{sup_f} * 100
\]


\item[Identity Error ($\sigma$)]
This metric reflects the percentage error in identifying frequent itemsets
and has two components: $\sigma^+$, indicating the percentage of false
positives, and $\sigma^-$ indicating the percentage of false
negatives.  Denoting the reconstructed set of frequent itemsets with $R$
and the correct set of frequent itemsets with $F$, these metrics are
computed as:
\begin{center}
${\sigma^+}  = \frac{\mid R - F\mid }{\mid F\mid } * 100$ \hspace*{0.5in}
${\sigma^-}  = \frac{\mid F - R\mid }{\mid F\mid }$ * 100
\end{center}
\end{description}

\paragraph{Perturbation Mechanisms.}
We show frequent-itemset-mining accuracy results for our proposed
perturbation mechanisms as well as representative prior techniques. For
all the perturbation mechanisms, the mining from the distorted database
was done using \emph{Apriori} \cite{as94} algorithm, with an additional
support reconstruction phase at the end of each pass to recover the
original supports from the perturbed database supports computed during
the pass \cite{mask,emask}.

The perturbation mechanisms evaluated in our study are the following:
\begin{description}

\item[DET-GD:]
This schemes uses the  deterministic gamma-diagonal perturbation matrix $A$
(Section \ref{sec:choiceA}) for
perturbation and reconstruction.  The implementation described in Section
\ref{sec:perturbation_algo} was used to carry out the perturbation,
and the results of Section \ref{sec:FI_mining} were used to compute the
perturbation matrix used in each pass of Apriori for reconstruction.

\item[RAN-GD:]
This scheme uses the randomized gamma-diagonal perturbation matrix
$\tilde{A}$ (Section \ref{sec:randomA}) for perturbation and reconstruction.
Though in general, any distribution can be used for $\tilde{A}$, here we
evaluate the performance of uniformly distributed $\tilde{A}$ given by
Equation~\ref{matrixA_bar}
over the entire range of the randomization parameter $\alpha$.

\item[MASK:]
This is the perturbation scheme proposed in \cite{mask}, which is intended
for boolean databases and is characterized by a single parameter $1-p$, which
determines the probability of an attribute value being flipped. 
In our scenario, the categorical attributes are mapped to boolean
attributes by making each value of the category an attribute. Thus,
the $M$ categorical attributes map to $M_b = \sum_{j}\mid S_U^j
\mid$ boolean attributes. 

The flipping probability $1-p$ was chosen as the lowest value which
could satisfy the constraints given by Equation \ref{pconstraint}.
The constraint 
$\forall v :\forall u_1,u_2 : \frac{A_{vu_1}}{A_{vu_2}} \leq \gamma $
is satisfied for MASK~\cite{mask}, if
$\displaystyle \frac{p^{M_b}}{(1-p)^{M_b}} \leq \gamma $.
But, for each categorical attribute, one and only one of its associated
boolean attributes takes value $1$ in a particular record.  Therefore,
all the records contain exactly $M$ $1^s$, and the following condition
is sufficient for the privacy constraints to be satisfied: 
\[\frac{p^{2M}}{(1-p)^{2M}} \leq \gamma \]. 
This equation was used to determine the appropriate value of $p$.
Value of $p$ turns out be $0.5610$ and $0.5524$ respectively for CENSUS and HEALTH datasets for $\gamma=19$.

\item[C\&P:]
This is the Cut-and-Paste perturbation scheme proposed in \cite{gehrke02}, with
algorithmic parameters $K$ and $\rho$.  To choose $K$, we varied $K$ from $0$
to $M$, and for each $K$, $\rho$ was chosen such that the matrix
(Equation~\ref{matrix_cut_paste}) satisfies the privacy constraints (Equation
\ref{pconstraint}). The results reported here are for the $(K,\rho)$
combination giving the best mining accuracy. For $\gamma=19$ $K=3, \rho=0.494$
turn out to be appropriate values.  

\shcomment{do we want to write this}
\end{description}

\begin{figure*}[tp]
\begin{center}
\subfigure[] {\includegraphics[width=2.1in,height=2in]{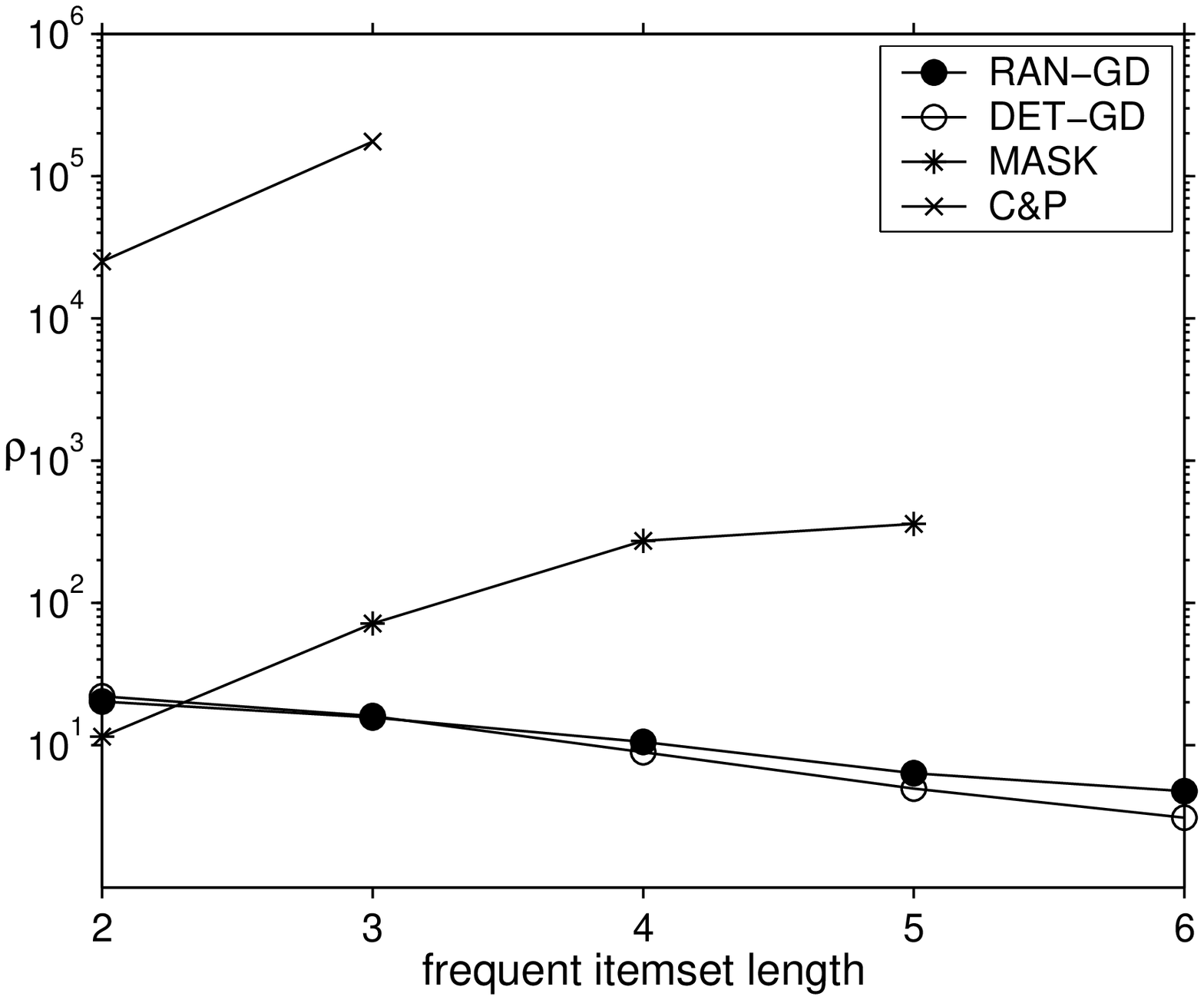}}
\hfill
\subfigure[] {\includegraphics[width=2.1in,height=2in]{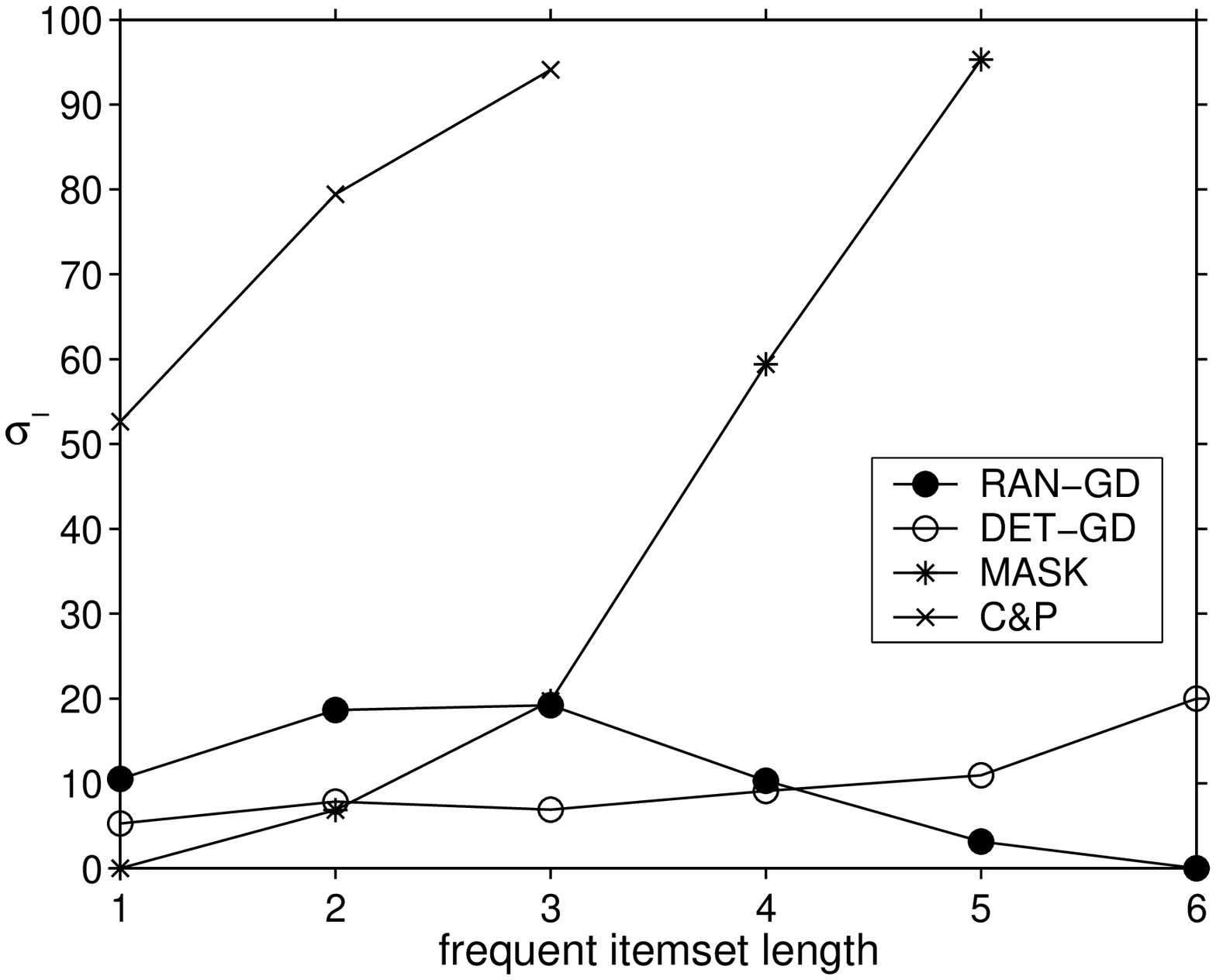}} 
\hfill
\subfigure[] {\includegraphics[width=2.1in,height=2in]{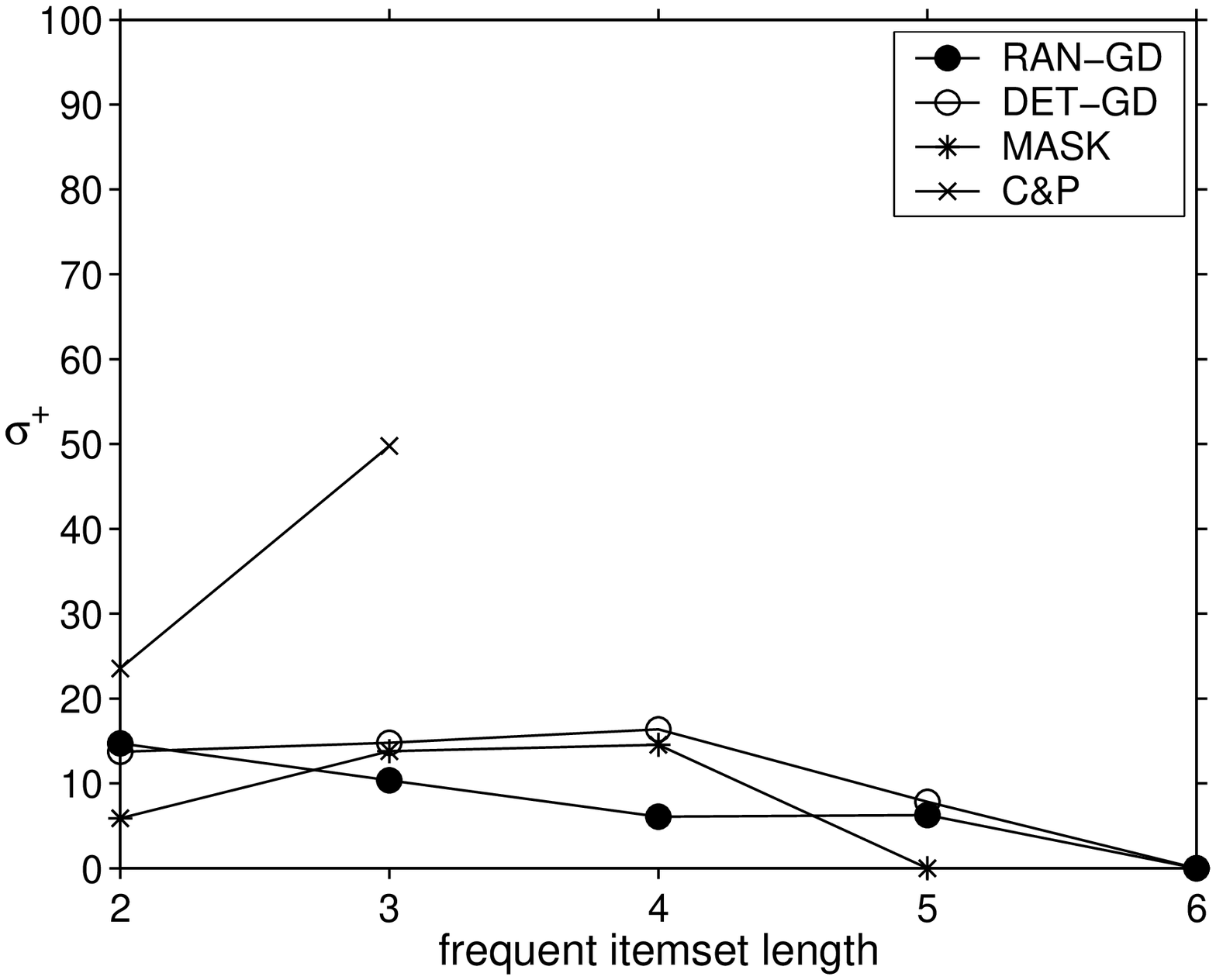}} 
\vspace{-0.6cm}
\caption{\small{(a) Support error $\rho$ (b) False negatives $\sigma^-$ (c) False positives $\sigma^+$ for CENSUS dataset}
\label{fig:adult_error}} 
\end{center}
\end{figure*}
\begin{figure*}
\begin{center}
\subfigure[] {\includegraphics[width=2.1in,height=2in]{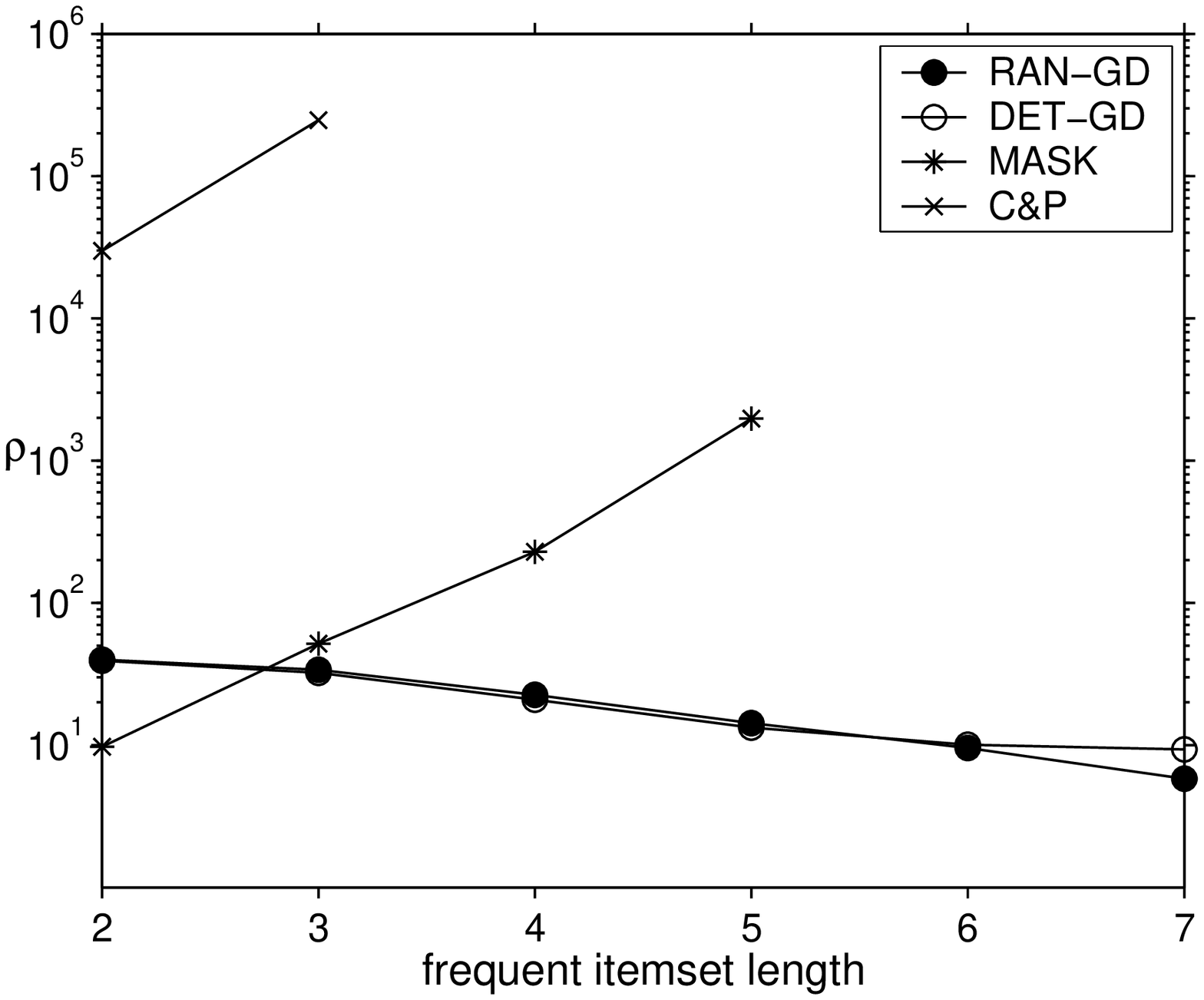}}
\hfill
\subfigure[] {\includegraphics[width=2.1in,height=2in]{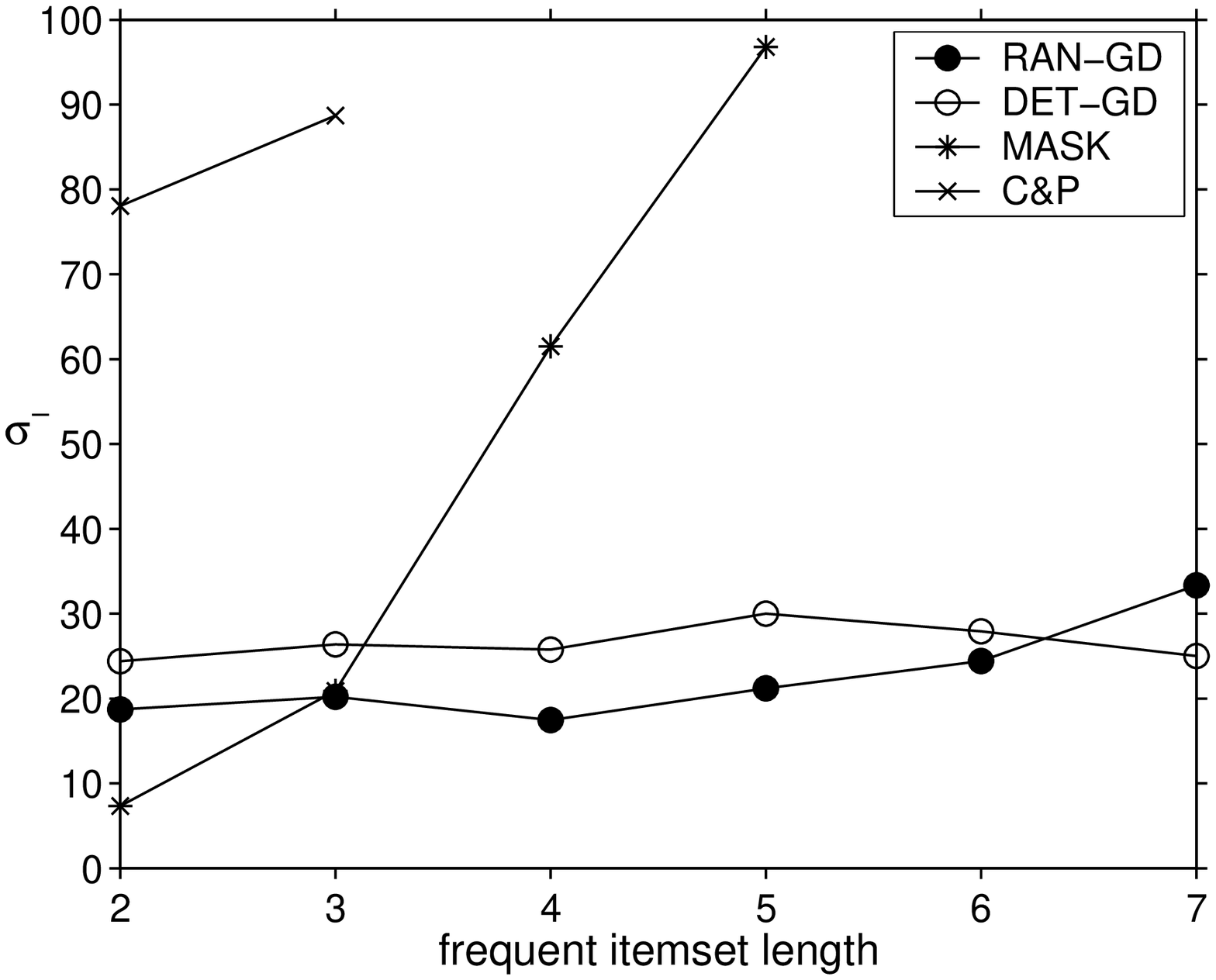}} 
\hfill
\subfigure[] {\includegraphics[width=2.1in,height=2in]{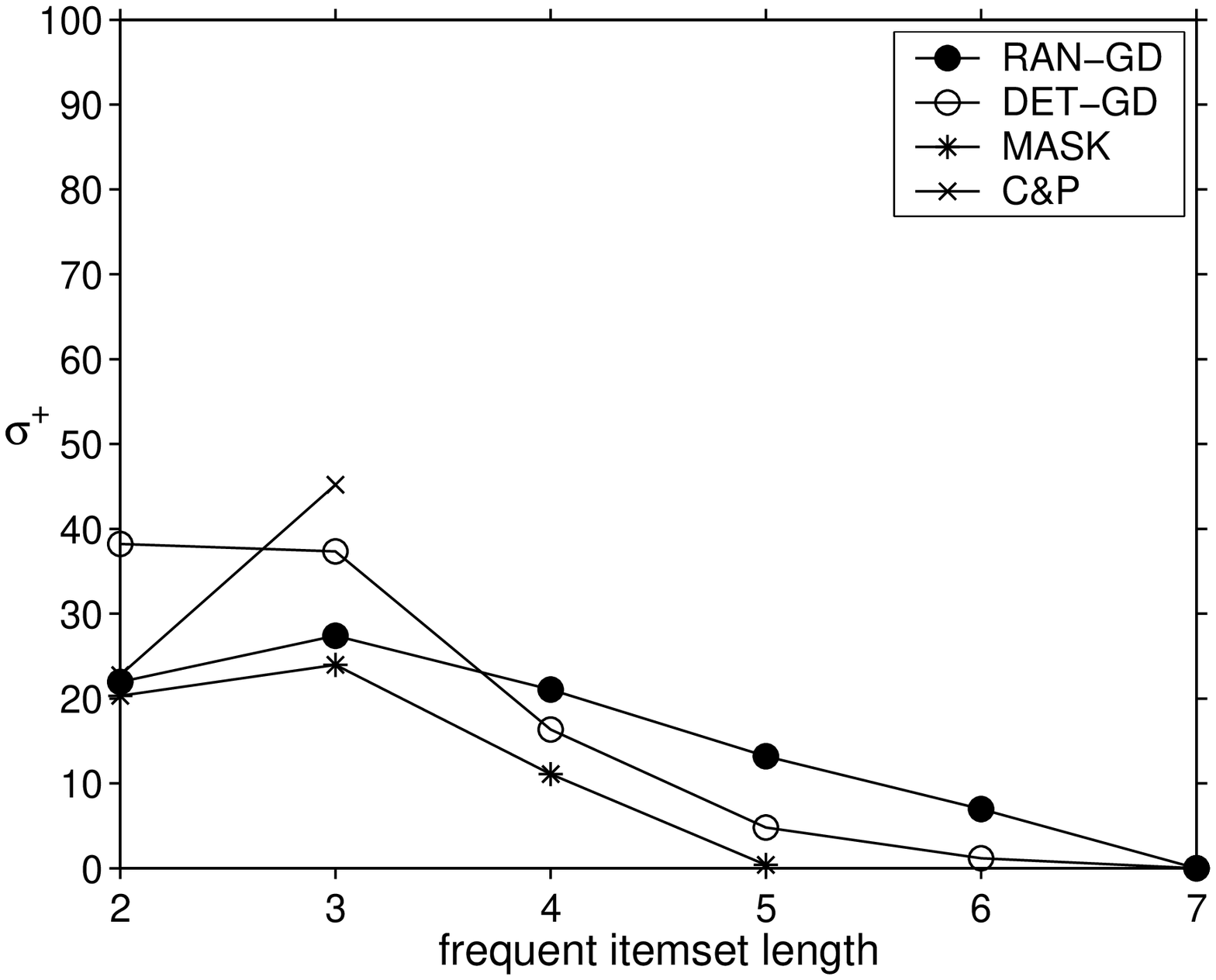}} 
\vspace{-0.6cm}
\caption{\small{(a) Support error $\rho$ (b) False negatives $\sigma^-$ (c) False positives $\sigma^+$ for HEALTH dataset}
\label{fig:health_error}} 
\end{center}
\end{figure*}

\SubSection{Experimental Results}
For the CENSUS dataset, the support ($\rho$) and identity ($\sigma^-$,
$\sigma^+$) errors of the four perturbation mechanisms (DET-GD, RAN-GD,
MASK, C\&P) is shown in Figure~\ref{fig:adult_error}, as a function of the
length of the frequent itemsets.  The corresponding graphs for the HEALTH
dataset are shown in Figure~\ref{fig:health_error}. In this graph for
comparison, the performance of RAN-GD is shown for randomization parameter
$\alpha={\gamma x}/{2}$. Note that the support error ($\rho$) is plotted on a
\emph{log-scale}.

In these figures, we first note that the performance of the DET-GD
method is visibly better than that of MASK and C\&P.  In fact, as the
length of the frequent itemset increases, the performance of both MASK and C\&P degrades
drastically. MASK is not able to find any itemsets of length above
4 for the CENSUS dataset, and above 5 for the HEALTH dataset, while C\&P does not works after 3-length itemsets. 

\shcomment{\small{such comments need to be added for cut-paste operator, I will add 
them once i get the final numbers}}

\begin{figure*}[tp]
\begin{center}
\subfigure[] {\includegraphics[width=2.1in,height=2in]{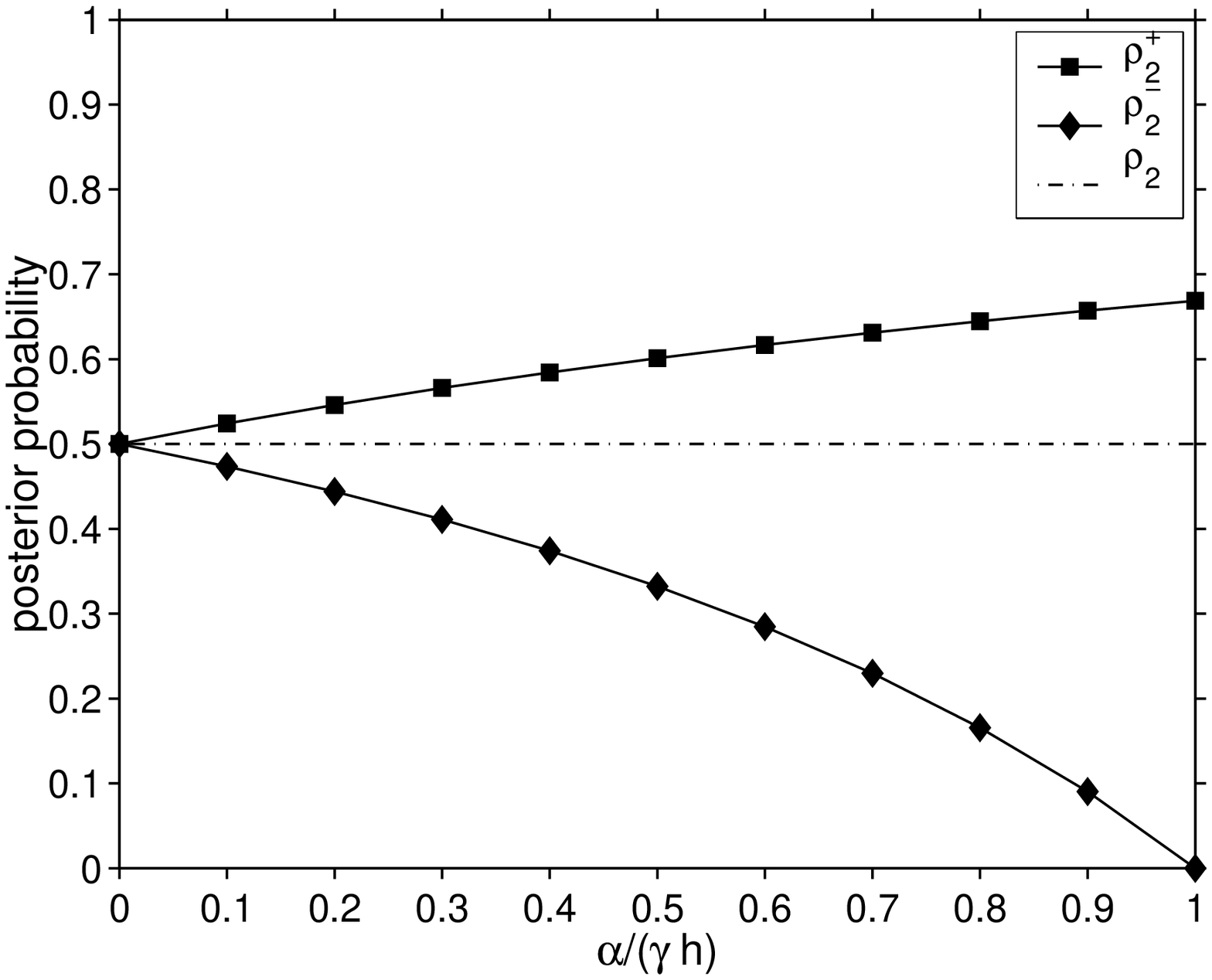}}
\hfill
\subfigure[] {\includegraphics[width=2.1in,height=2in]{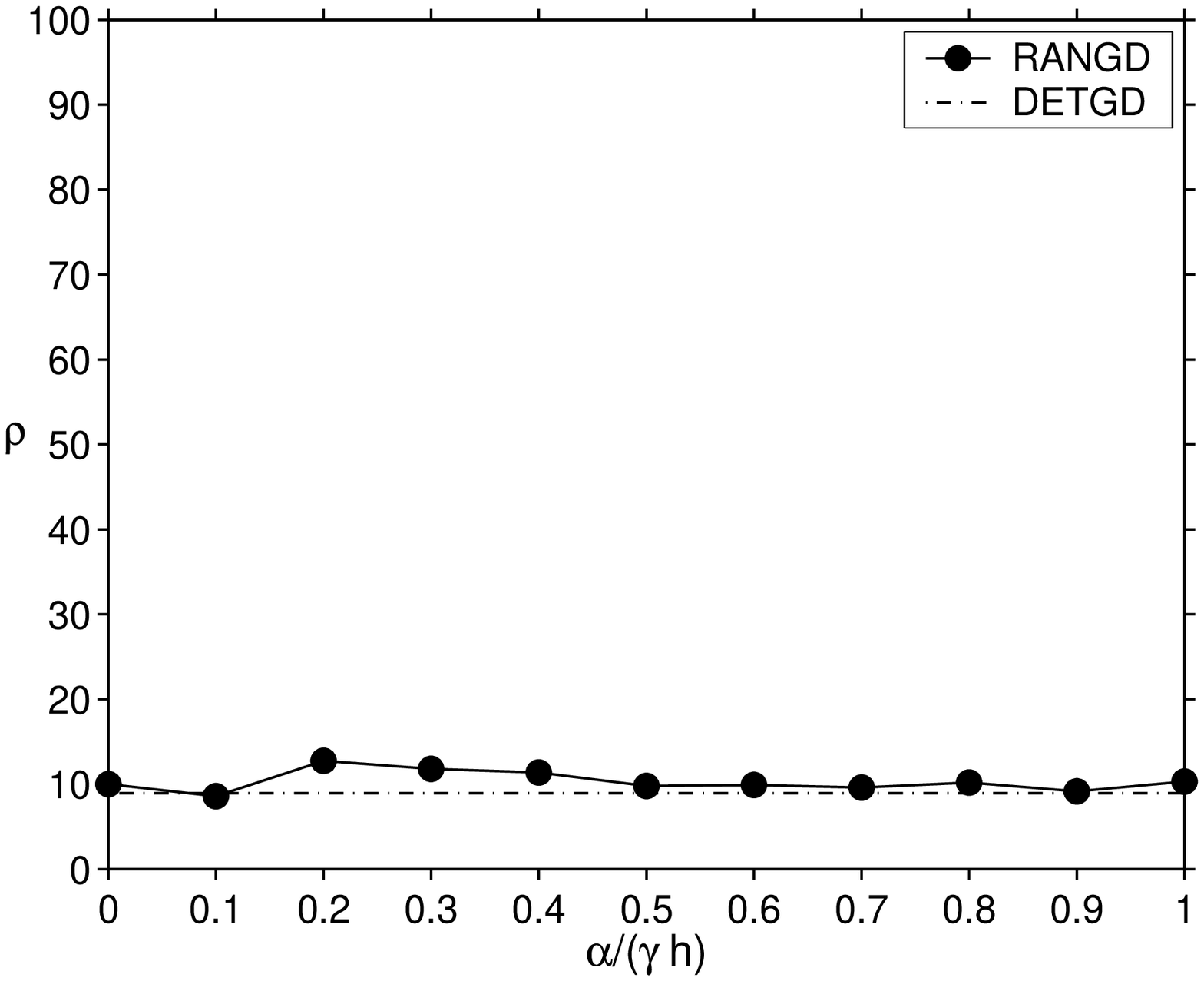}}
\hfill
\subfigure[] {\includegraphics[width=2.1in,height=2in]{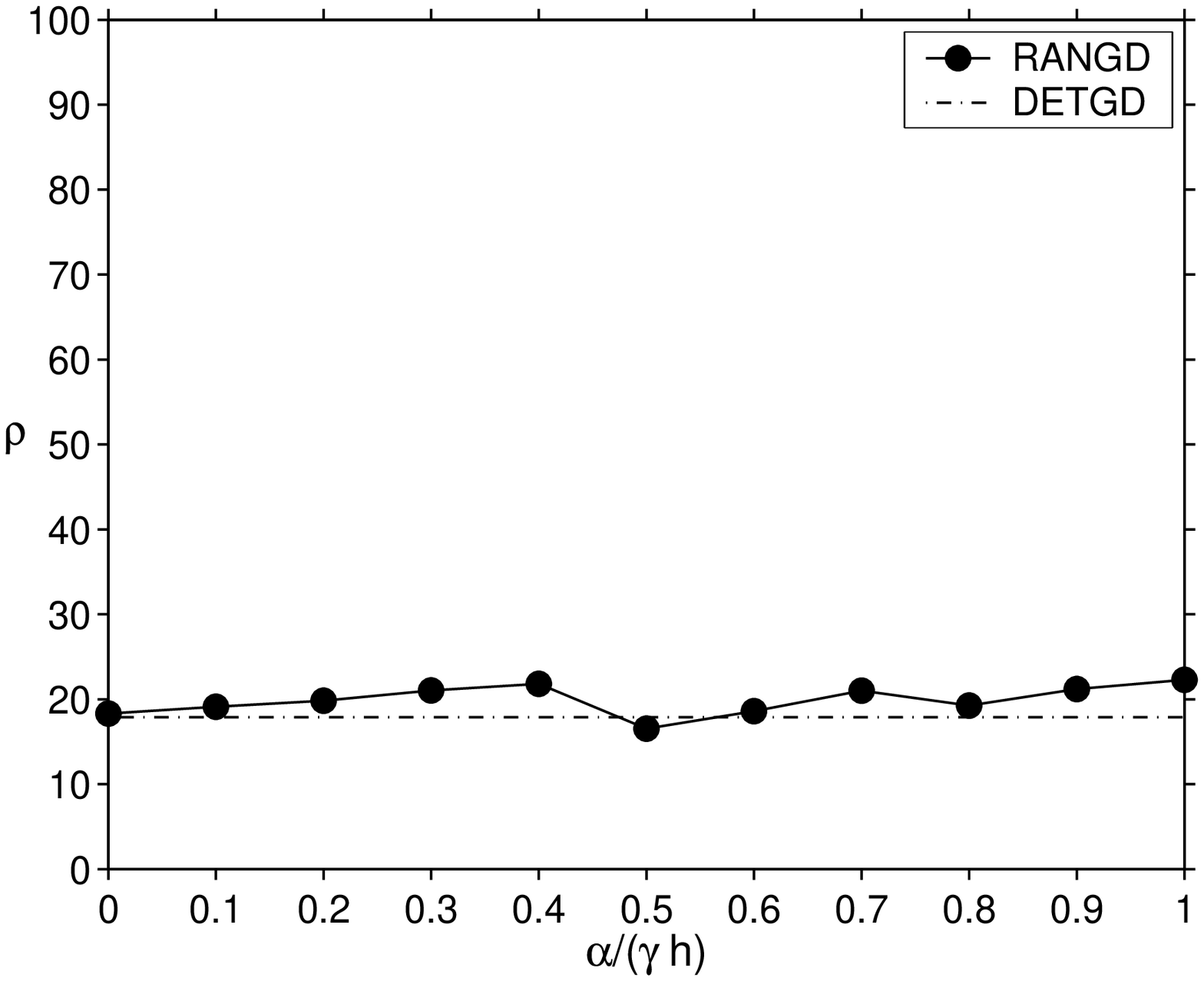}}
\vspace{-0.6cm}
\caption{\small{(a) Posterior probability ranges (b) Support error $\rho$ for CENSUS (c) Support error $\rho$ for HEALTH dataset with varying degree of randomization }
\label{fig:rangd}} 
\end{center}
\end{figure*}

The second point to note is that the accuracy of RAN-GD, although dealing
with a randomized matrix, is only marginally lower than that of DET-GD.
\shcomment{it is actually not lower it varies, sometimes higher and sometimes lower}
In return, it provides a substantial increase in the privacy -- its worst
case (determinable) privacy breach is only $33\%$ as compared to
$50\%$ with DET-GD. Figure~\ref{fig:rangd} 
shows performance of RAN-GD over entire range of $\alpha$, and the posterior probability range $[\rho^-,\rho^+]$. It shows mining support reconstruction errors for itemset length $4$. We can observe that the performance of RAN-GD does not deviate much from the derterministic case over the entire range, where as very low \emph{determinable} posterior probability can be obtained for higher values of $\alpha$. 

\shcomment{write inferences from the figure \ref{fig:adult_rangd} and \ref{fig:health_rangd}}

\begin{figure*}
\begin{center}
\hspace{1in}
\subfigure[] {\includegraphics[width=2.1in,height=2in]{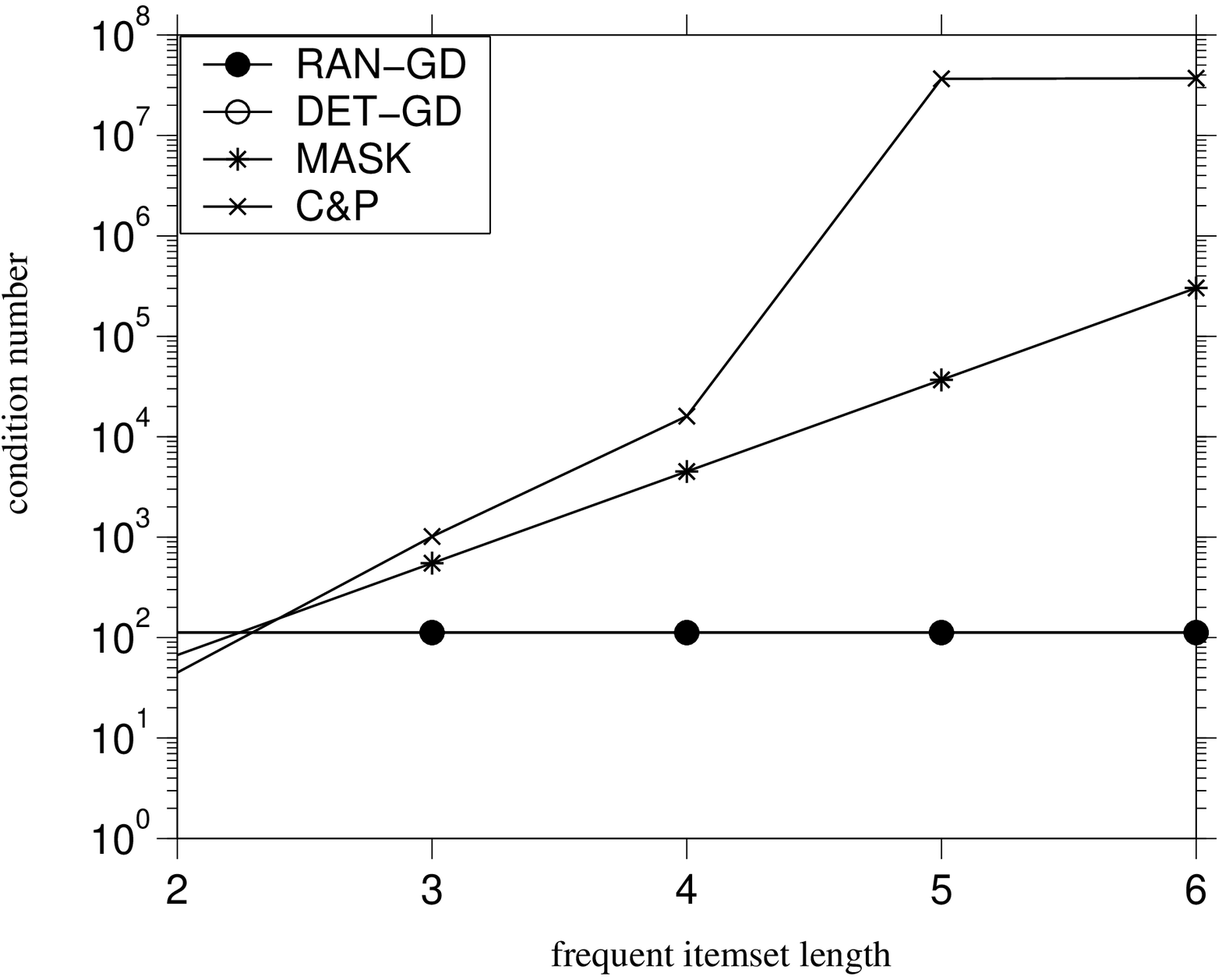}}
\hfill
\hspace{0.5cm} 
\subfigure[] {\includegraphics[width=2.1in,height=2in]{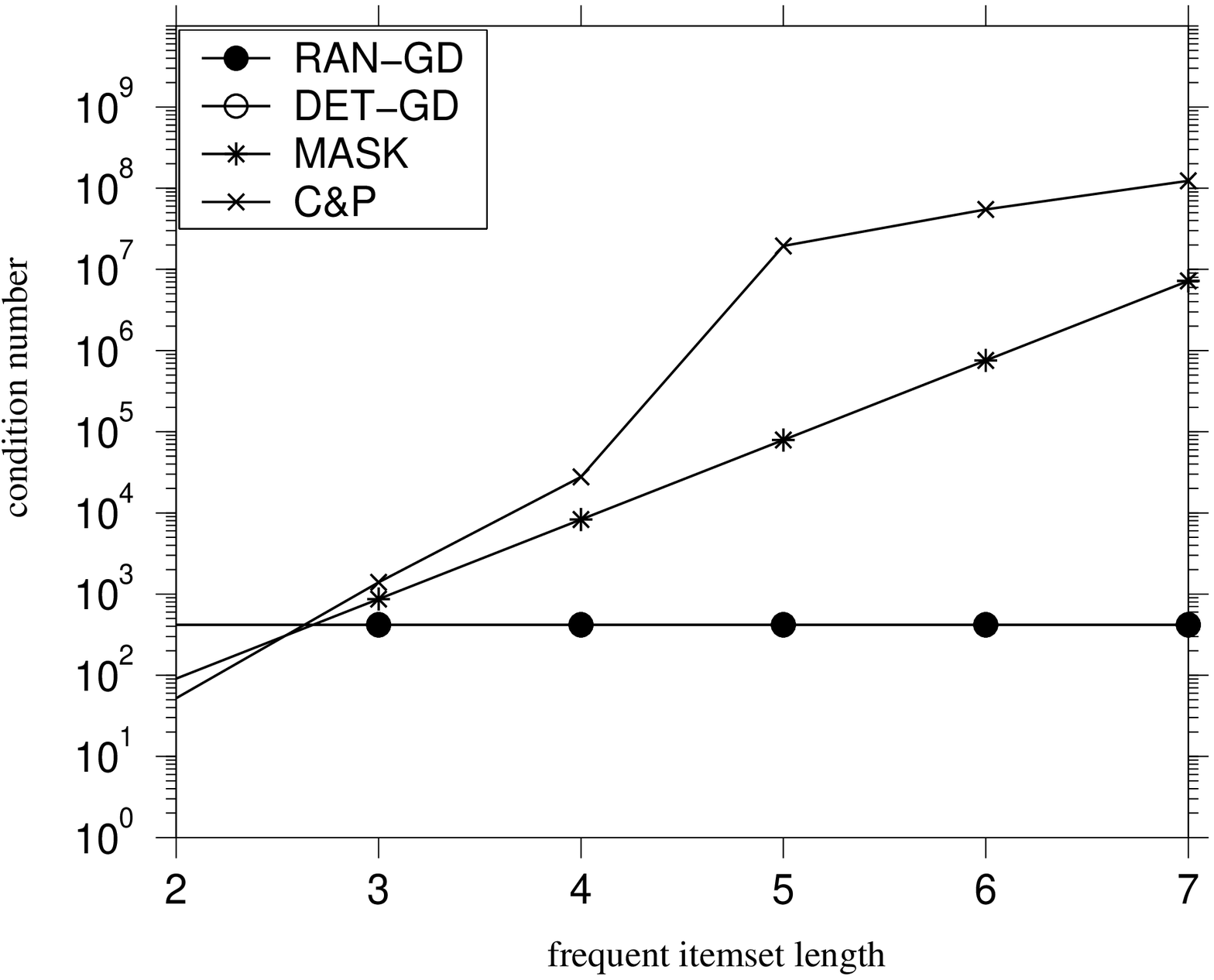}} 
\hspace{1in}
\hfill
\vspace{-0.6cm}
\caption{\small{Comparison of condition number of transition probability matrix (a) CENSUS (b) HEALTH }
\label{fig:cond_number}} 
\end{center}
\end{figure*}

The primary reason for DET-GD and RAN-GD's good performance is the
low \emph{condition number} of their perturbation matrices. This is
quantitatively shown in Figure~\ref{fig:cond_number}, which compares
the condition numbers (on a \emph{log-scale}) of the reconstruction
matrices. Note that as the expected value of random matrix $\tilde{A}$ is
used for estimation in RAN-GD, and the random matrix used in experiments
has expected value $A$ (refer Equation~\ref{matrixA_bar}) used in DET-GD,
the condition numbers for two methods are equal. Here we see that the
condition number for DET-GD and RAN-GD is not only low but also constant
over all lengths of frequent itemsets (as mentioned before, the condition
number is equal to $1 + \frac{|S_U|)}{(\gamma-1)}$). In marked contrast,
the condition number for MASK and C\&P increase \emph{exponentially} with
increasing itemset length, resulting in drastic degradation in accuracy.
Thus our choice of a gamma-diagonal matrix shows highly promising
results for discovery of long patterns.

\jhcomment{What is the condition number formula for RAN-GD?}

\comment{
\begin{figure*}[tp]
\begin{center}
\subfigure[] {\includegraphics[width=2.1in,height=2in]{Graphs/adult_privacy_aratio.eps}}
\hfill
\subfigure[] {\includegraphics[width=2.1in,height=2in]{Graphs/adult_rho_aratio.eps}}
\hfill
\subfigure[] {\includegraphics[width=2.1in,height=2in]{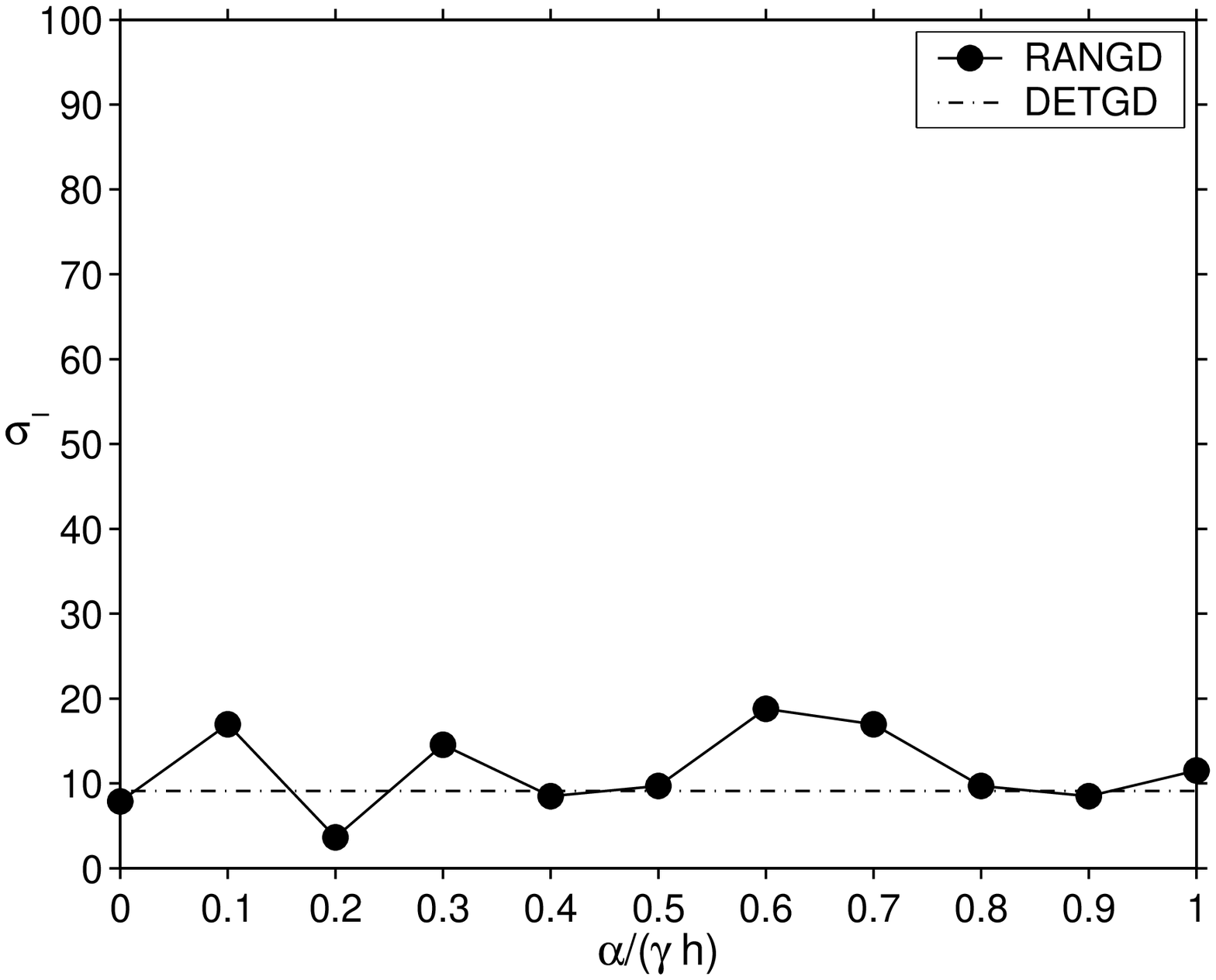}} 
\hfill
\subfigure[] {\includegraphics[width=2.1in,height=2in]{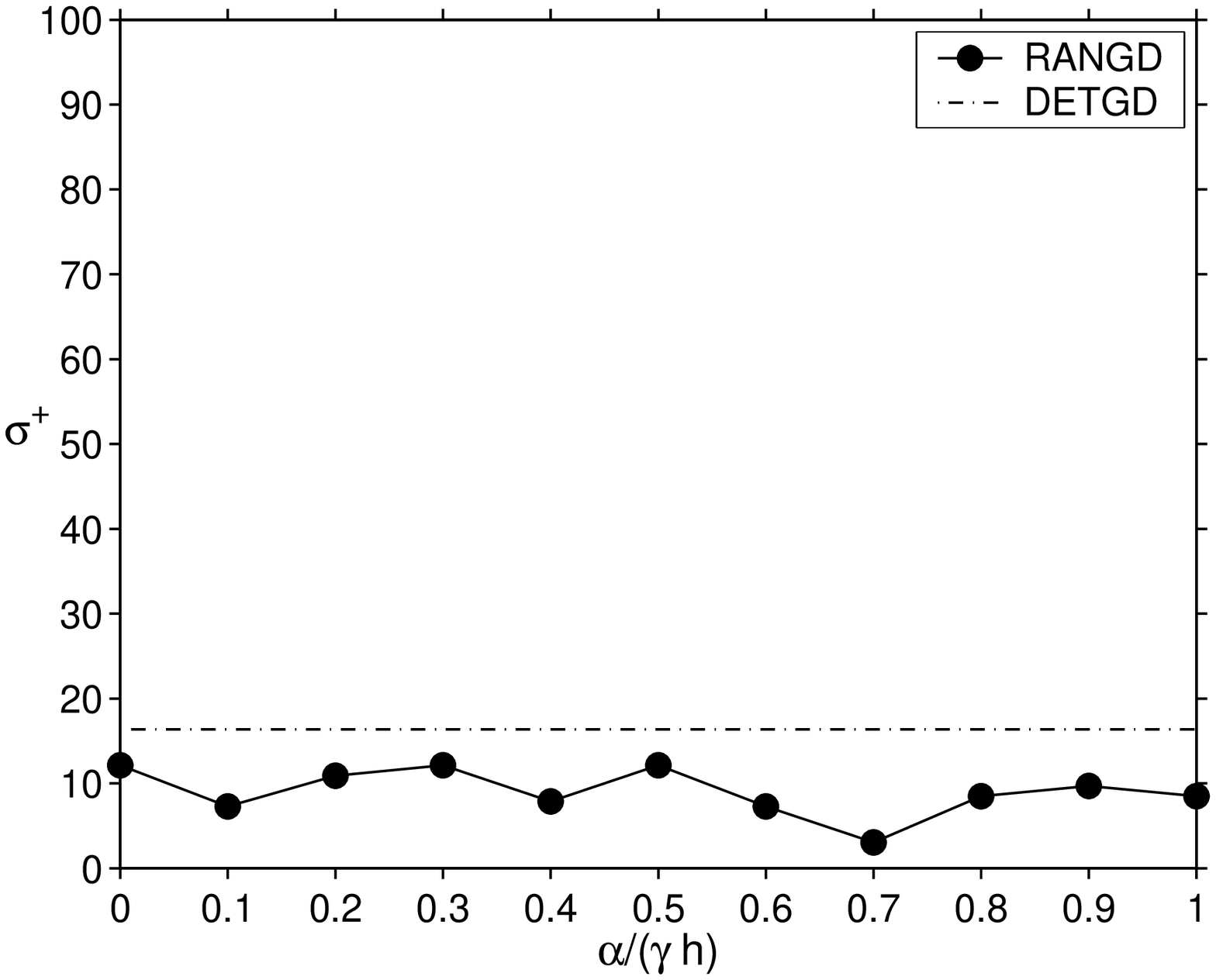}} 
\vspace{-0.6cm}
\caption{\small{(a)posterior probability ranges (b) Support error $\rho$ (c) False negatives $\sigma^-$ (d) False positives $\sigma^+$ for itemset length 4 by RAN-GD with varying degree of randomization (CENSUS dataset) }
\label{fig:adult_rangd}} 
\end{center}
\end{figure*}

\begin{figure*}[tp]
\begin{center}
\subfigure[] {\includegraphics[width=2.1in,height=2in]{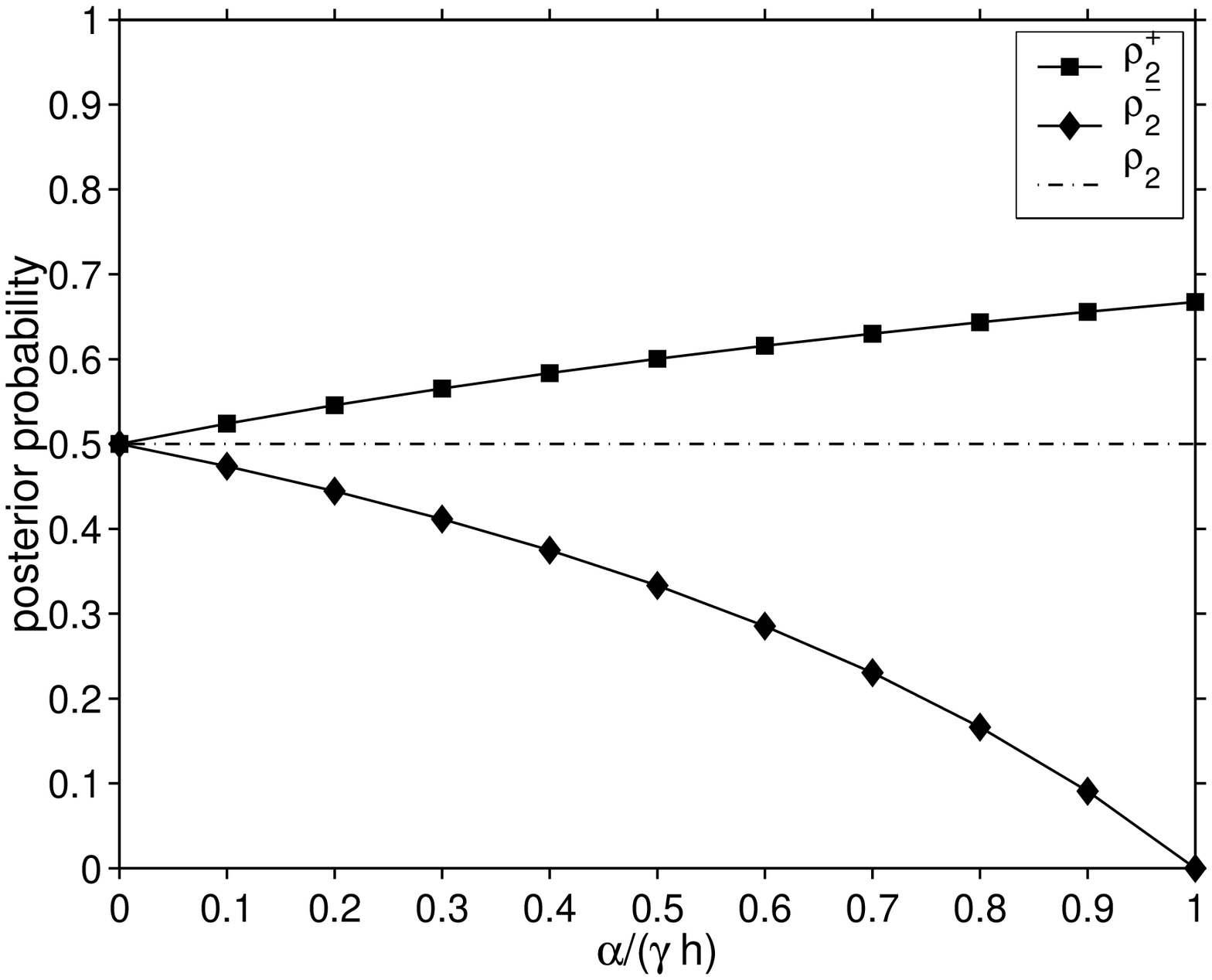}}
\hfill
\subfigure[] {\includegraphics[width=2.1in,height=2in]{Graphs/health_rho_aratio.eps}}
\hfill
\subfigure[] {\includegraphics[width=2.1in,height=2in]{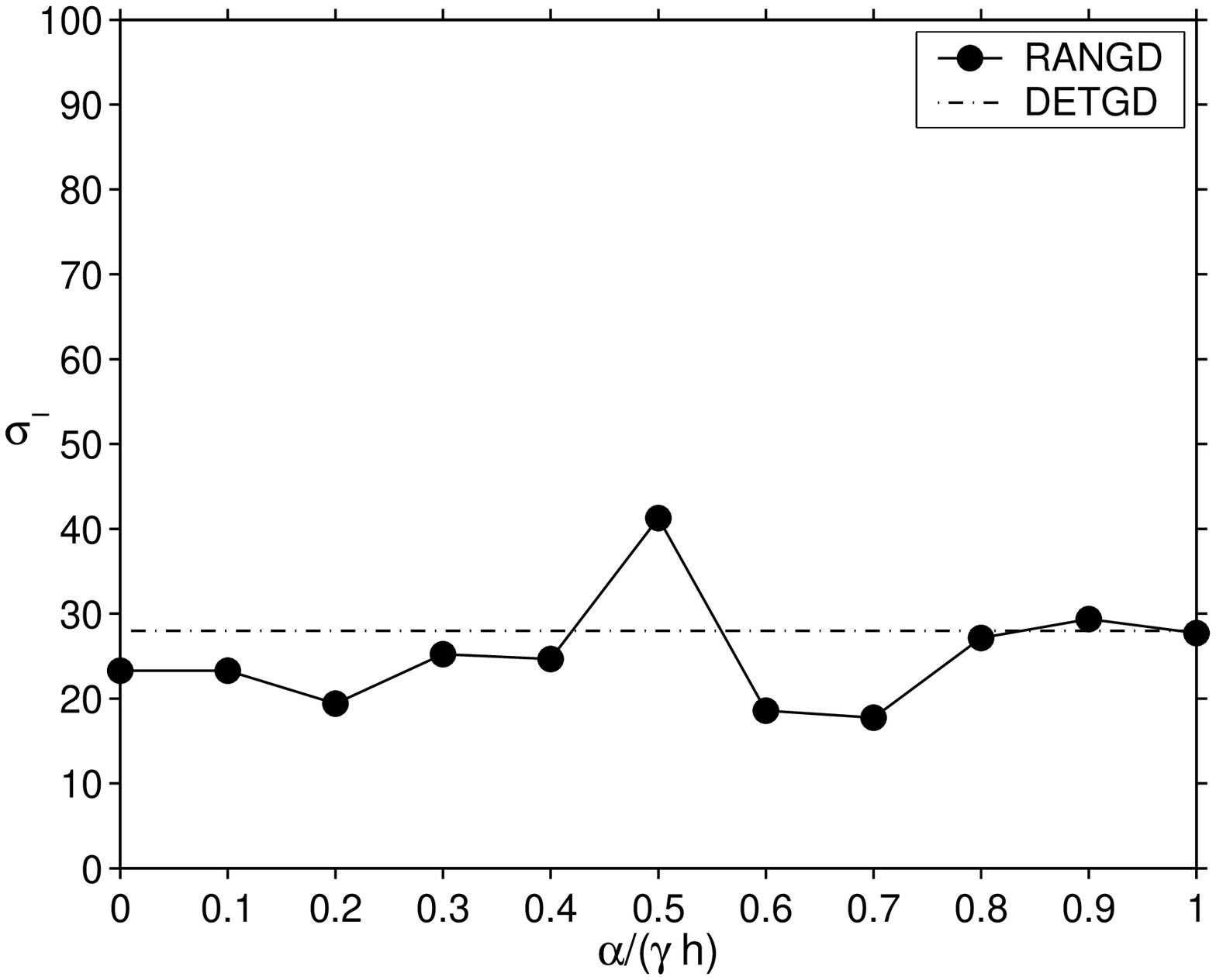}} 
\hfill
\subfigure[] {\includegraphics[width=2.1in,height=2in]{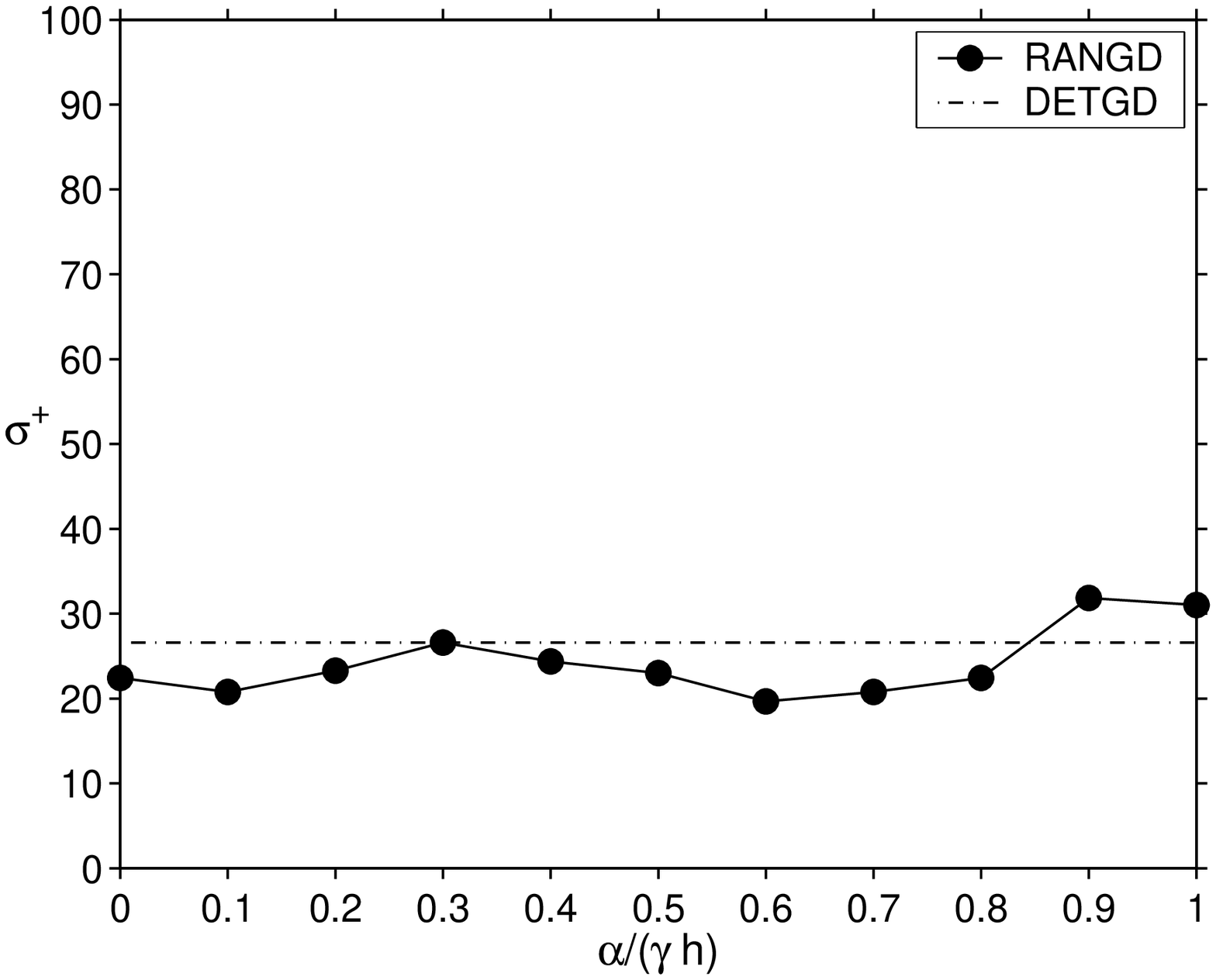}} 
\vspace{-0.6cm}
\caption{\small{(a)posterior probability ranges (b) Support error $\rho$ (c) False negatives $\sigma^-$ (d) False positives $\sigma^+$ for itemset length 4 by RAN-GD with varying degree of randomization (HEALTH dataset) }
\label{fig:health_rangd}} 
\end{center}
\end{figure*}
}

\Section{Related Work}
\label{sec:related}

The issue of maintaining privacy in data mining
has attracted considerable attention in the recent past.
\comment{
(e.g.~\cite{very01,very01a,very01b,very01c,vaidya,clif02,gehrke02,mask,lim03,as00,charu01}).
}

\jhcomment{give the references in increasing number.}

The work closest to our approach is that of
\cite{as00,charu01,gehrke02,mask,lim03}. In the pioneering work of
\cite{as00}, privacy-preserving data classifiers based on adding noise
to the record values were proposed.  This work was extended in
\cite{charu01} and \cite{kargupta03} to address a variety of
subtle privacy loopholes.

\comment{
examined the problem
of quantifying privacy and pointed out the potential of privacy loss in
such perturbation if the domain of original value was different from the
domain of perturbed values.  A signal processing approach was used by
\cite{kargupta03} to show that that perturbation through additive noise
is liable for data value reconstruction through filtering if the signal
to noise ratio is not very large.
}

New randomization operators for maintaining data privacy for boolean data
were presented and analyzed in \cite{gehrke02,mask}. These methods are
for categorical/boolean data and are based on probabilistic mapping from
domain space to the range space rather than by incorporating additive
noise to continuous valued data.  A theoretical formulation of privacy breaches
for such methods and a methodology for limiting them were given in the
foundational work of \cite{lim03}.

Our work is directly related to the above-mentioned methodologies
for privacy preserving mining. We combine the approaches for random
perturbation on categorical data into a common theoretical framework,
and explore how well random perturbation methods can do in the face of
strict privacy requirements. We show that we can derive a perturbation
matrix which performs significantly better than the existing methods for
discovery of frequent itemsets in categorical data while simultaneously
ensuring strict privacy guarantees. Also, we propose the novel idea of
making the perturbation matrix itself random which, to the best of our
knowledge, has not been previously explored in the context of privacy
preserving mining.


Another model of privacy preserving data mining is k-anonymity model
\cite{k-anony}. The perturbation approach used in random perturbation model
works under the strong privacy requirement that even the dataset forming server
is not allowed to learn or recover precise records. Users trust nobody and
perturb their record at their end before providing it to any other party.
k-anonymity model\cite{k-anony} does not satisfy this requirements. The condensation approach
discussed in \cite{condensation} also requires the relaxation of the assumption
that even the data forming server is not allowed to learn or recover records,
as in k-anonymity model. Hence these models are orthogonal to our privacy model.

\cite{hipp02,hipp04a,hipp04b,hipp04c} deal with Hippocratic databases which are
the database systems that take responsibility of the privacy of data they manage. It
involves specification of how the data is to be used in a privacy policy and
enforcing limited disclosure rules for regulatory concerns prompted by
legislation.

Finally, the problem addressed in \cite{very01,very01a,very01b,very01c} is
how to prevent \emph{sensitive rules} from being inferred by the
data miner -- this work is complementary to ours since it addresses
concerns about \emph{output} privacy, whereas our focus is on the
privacy of the \emph{input} data.  Maintaining input data privacy is
considered in \cite{vaidya,clif02,clif03,clif04} in the context of databases that are
\emph{distributed} across a number of sites with each site only willing
to share data mining results, but not the source data.

\Section{Conclusions and Future Work}
\label{sec:conc}
\shcomment{\small{Please use 'FRAPP' framework wherever required in this part}}
In this paper, we developed FRAPP, a generalized model for
random-perturbation-based methods operating on categorical data under
strict privacy constraints.  We showed that by making careful choices of
the model parameters and building perturbation methods for these choices,
order-of-magnitude improvements in  accuracy could be achieved as compared
to the conventional approach of first deciding on a method and thereby
implicitly fixing the associated model parameters.  In particular,
we proved that a ``gamma-diagonal'' perturbation matrix is capable
of delivering the best accuracy, and is in fact, optimal with respect
to its condition number.  We presented an implementation technique for
gamma-diagonal-based perturbation, whose complexity is proportional to the
\emph{sum} of the domain cardinalities of the attributes in the database.
Empirical evaluation of our new gamma-diagonal-based techniques on
the CENSUS and HEALTH datasets showed substantial reductions in frequent
itemset identity and support reconstruction errors.

We also investigated the novel strategy of having the perturbation
matrix composed of not values, but random variables instead.  Our
analysis of this approach showed that, at a marginal cost in accuracy,
signficant improvements in privacy levels could be achieved.

In our future work, we plan to extend our modeling approach to other
flavors of mining tasks.


\shcomment{\small{should we add the task of finding the minimum condition number matrix also as an open problem?}}


\bibliographystyle{latex8}

\baselineskip 10pt

\end{document}